\documentclass[aps,prd,floatfix,showpacs,showkeys,superscriptadress,unsortedaddress,nofootinbib,onecolumn]{revtex4-1}
\usepackage{mathtools,amsmath,amssymb,bbm,bm,slashed}
\usepackage{ulem}
\usepackage{graphicx,float}
\usepackage{hyperref}
\usepackage{color,bbold}
\hypersetup{colorlinks=true,linkcolor=blue,citecolor=blue,urlcolor=blue}
\usepackage{morefloats,subfigure}
\usepackage[toc,page]{appendix}
\usepackage{xcolor}

\newcommand{\sF}{\scriptscriptstyle{f}}
\newcommand{\sB}{\scriptscriptstyle{(B)}}
\newcommand{\sV}{\scriptscriptstyle{V}}
\newcommand{\sperp}{\scriptscriptstyle{\perp}}
\newcommand{\wt}[1]{\widetilde{#1}}
\newcommand{\shp}{\shortparallel}
\newcommand{\be}{\begin{eqnarray}}
\newcommand{\ee}{\end{eqnarray}}
\newcommand{\nn}{\nonumber \\}

\begin{document}
\title{Lepton pair production from a hot and dense QCD medium in the presence of an arbitrary magnetic field}
\author{Aritra Das$^{1,2,a}$, Aritra Bandyopadhyay$^{3,4,b}$, Chowdhury Aminul Islam$^{5,c}$}
\affiliation{$^1$ HENPP Division, Saha Institute of Nuclear Physics, HBNI, 1/AF Bidhan Nagar, Kolkata 700064, India \\
$^2$ School of Physical Sciences, National Institute of Science Education and Research, HBNI, Jatni, Khurda 752050, India}
\affiliation{$^3$Guangdong Provincial Key Laboratory of Nuclear Science, Institute of Quantum Matter, South China Normal University, Guangzhou 510006, China}
\affiliation{$^4$Institut für Theoretische Physik, Universität Heidelberg, Philosophenweg 16, 69120 Heidelberg, Germany}
\affiliation{$^5$ School of Nuclear Science and Technology, University of Chinese Academy of Sciences, Beijing 100049, China}
\email{$^a$ aritra.das@niser.ac.in}
\email{$^b$ a.bandyopadhyay@thphys.uni-heidelberg.de}
\email{$^c$ chowdhury.aminulislam@gmail.com}

\begin{abstract}
{In this article, we have explored the very important quantity of lepton pair production from a hot and dense QCD medium in presence of an arbitrary external magnetic field for simultaneous nonzero values of both the parallel (along the direction of the external field) and perpendicular (lying on the transverse plane to the external field) components of the dilepton momentum. As opposed to the zero magnetic field case (the so-called Born rate) or the lowest Landau level approximated rate, where only the annihilation process contributes, here we observe contributions also arising out of the quark and antiquark decay processes. We found the encouraging result of considerable enhancement of lepton pair production in presence of an arbitrary magnetic field. We decompose the total rate into different physical processes and discuss their behaviors for both zero and nonzero baryon density. The whole analysis is then subjected to an effective model treatment, where we have incorporated the magnetic field induced novel effects of magnetic catalysis (MC) and inverse MC (IMC) through a medium dependent scalar coupling, which leads to some further interesting observations.}
\end{abstract}

\maketitle

\section{Introduction}
\label{sec:intro}
Studying many particle systems, particularly that of the strongly interacting quantum chromodynamics (QCD) medium created in heavy ion collisions (HIC) is one of the most interesting areas of high energy physics~\cite{Muller:1983ed,Heinz:2000bk,Pasechnik:2016wkt}. The richness of the subject is constantly growing with time, specifically due to the gradual incorporation of different extreme conditions that the created medium can be subjected to in such collisions. These extreme conditions can be in terms of tremendously high temperature, density, magnetic field, isospin density etc~\cite{Fukushima:2011jc,Yagi:2005yb}.

A lot of efforts have been put to understand the properties of the medium in presence of a magnetic field ($eB$), which is supposed to be robust in magnitude in HICs because of their noncentrality \textemdash\, $eB\sim m_\pi^2$ in Relativistic Heavy Ion Collider (RHIC) and $\sim10\,m_\pi^2$ in Large Hadron Collider (LHC)~\cite{Skokov:2009qp}. Thus, it is expected that the medium properties will be affected by the magnetic field~\cite{Kharzeev:2013jha,Miransky:2015ava}. The same has been demonstrated by several interesting studies in the recent past, in the process catering some novel effects, e.g. magnetic catalysis (MC)~\cite{Gusynin:1994re}, inverse magnetic catalysis (IMC)~\cite{Bali:2012zg,Bali:2011qj}, chiral magnetic effect (CME)~\cite{Fukushima:2008xe} etc. One may also assume the strong magnetic field created in such collisions to be present for an extended period during the evolution of the QCD medium starting from the noncentral HICs~\cite{Tuchin:2015oka, Skokov:2009qp,Guo:2019mgh}. Although recently some opposing views have been proposed which suggest that the search for any magnetic effects in the HIC would be highly challenging due to the short lifetime of the induced magnetic field~\cite{Wang:2021oqq}, in accordance with the non-observation of CME signals in the isobar experiments~\cite{STAR:2021mii}

It is a well-known and well-explored fact that the properties of such many particle systems in extreme conditions are embedded in the correlation function (CF) and its spectral representation~\cite{Forster:1975pm,Callen:1951vq,Kubo:1957mj}. Now, while propagating through the magnetised hot and/or dense medium the vacuum properties of any particle is bound to be affected, which in turn modifies the spatial and temporal CFs. Hence studying the structure of the correlation function and the corresponding spectral representation are of immense importance. For example the vector-vector current CF is associated with the differential thermal cross section for the dilepton production rate (DR)~\cite{Weldon:1990iw}. The estimation of DR is of great importance, particularly because of the lepton pair leaving the fire ball with minimum interaction and thus carrying less contaminated information of the stage when it is created. Also, considering the short lifetime of the magnetic field produced in HIC, DR, being a relatively early observable, becomes an even more relevant signature of the magnetized medium. DR is a static quantity which acts as an input in the space-time evolution to obtain the dilepton spectra, the corresponding dynamic quantity which can be compared with the experimental data.

There are many instances where this kind of investigation has been performed both in absence and presence of magnetic field. For example, the vector-vector current CF and from there the DR were calculated in the effective model scenario in absence of magnetic field~\cite{Islam:2014sea}. Similar studies were also done but using a different model in the Refs.~\cite{Gale:2014dfa,Hidaka:2015ima}. The DR in presence of an extreme magnetic field was, possibly, first addressed in Refs.~\cite{Tuchin:2012mf,Tuchin:2013bda,Tuchin:2013ie} in a phenomenological manner. Using formal field theoretic approach the DR was estimated for a magnetised hot and dense medium using Ritus eigenfunction method in~\cite{Sadooghi:2016jyf,Hattori:2020htm} and using Schwinger proper time method in the strong field limit in imaginary time formalism~\cite{Bandyopadhyay:2016fyd} and in the weak field limit in real time formalism~\cite{Bandyopadhyay:2017raf}. It was revisited in the latter formalism for arbitrary magnetic field with the component of dilepton momentum perpendicular to the direction of magnetic field ($p_{\sperp}$) being zero and an enhancement was observed in the low mass region~\cite{Ghosh:2018xhh}. In another study~\cite{Islam:2018sog} the effect of magnetic field was incorporated through the effective mass, which was calculated in effective models like NJL and PNJL, and then spectral function and spectral properties like DR were estimated in the strong field limit. Recently similar studies have been performed for arbitrary magnetic field with $p_{\sperp}=0$ in the real time formalism, where the effect of anomalous magnetic moment has been incorporated~\cite{Ghosh:2020xwp}. In another recent study the photon self energy has been calculated in arbitrary magnetic field in the imaginary time formalism~\cite{Wang:2020dsr}. The authors then used that to estimate the direct photon rate and there-from the ellipticity of photon emission. The details of the calculation of photon self energy along with the estimation of magneto-optical conductivity is given in~\cite{Wang:2021ebh}.

As argued above the calculation of DR in presence of an arbitrary magnetic field is of great importance. Previously, such an attempt had been made using the Ritus eigenfunction method~\cite{Sadooghi:2016jyf} where the authors found a novel anisotropy between the longitudinal and transverse parts of the DR. In the present manuscript we evaluate the DR in an imaginary time field theoretic approach using the Schwinger method~\cite{Schwinger:1951nm}. Similar attempt has also been made in real time field theory but with the approximation of $p_{\sperp}$ being always zero~\cite{Ghosh:2018xhh}. In terms of the calculation we make novel estimate of the rate for arbitrary values of $p_{\sperp}$, thereby extending the spectrum of validity of the DR through our results. Decomposing our total DR into its constituent physical processes we found that the dilepton production is dominated by the quark/antiquark decay and quark-antiquark annihilation processes in the lower and higher values of invariant mass, respectively. Interestingly, we observe an enhancement of DR as compared to the Born rate for relatively higher values of invariant mass particularly for high values of magnetic fields relevant to HICs, which was restricted to only low invariant mass in the previous study~\cite{Ghosh:2018xhh}. Recently, there is a study of DR with similar formalism as us, which
expresses the DR in terms of the transverse momentum lying in the collisional plane, rapidity and azimuthal angle,
and mainly focuses on the angular dependence of the DR along with its ellipticity~\cite{Wang:2022jxx}.

As the lepton pairs are produced in a magnetised medium, one needs to ponder over the question whether they get affected by the magnetic field along with the initial quarks. It appears to be obvious that they get affected by the field and that is what has been assumed in refs.~\cite{Sadooghi:2016jyf} and \cite{Ghosh:2018xhh}, where the authors have considered magnetized lepton pairs which makes their calculation more involved, but physically not adding much to the result. The reason of that being the larger mean free path of the leptons as compared to the plasma fireballs. Subsequently, the leptons come out of the fireball without much interaction and acquire a well-defined four momenta before reaching the detectors, as argued in the recent calculation~\cite{Wang:2022jxx}. Following this argument, in the present calculation, the effect of the magnetic field has only been considered for the initial quark-antiquark pairs.

As another major feat of the present study, we have also looked for the consequences of a more practical scenario, i.e., quarks treated as quasiparticles within an effective model like Nambu\textemdash Jona-Lasinio (NJL)~\cite{Farias:2014eca,Farias:2016gmy}. We have tuned the model to incorporate both the well known MC at low temperatures and IMC effects around the transition temperature~\cite{Bali:2012zg,Bali:2011qj} through a medium dependent scalar coupling $G_S(eB,T)$~\cite{Farias:2016gmy}. This in turn impacts the DR through the temperature and magnetic field dependent effective quark mass arising therefrom. Particularly, to capture the essence of the IMC effect in the DR, we have minutely observed the behavior of the rate near the crossover temperature for different values of the magnetic field. We have also discussed the appearance of a mass gap at lower values of the temperature which refers to higher values of the effective quark mass. This gap shows some interesting features with respect to the $p_{\sperp}$. It widens as well as shifts towards left when the value of $p_{\sperp}$ is increased. On the other hand, the line connecting the gap becomes stronger as we up the strength of $p_{\sperp}$.

The paper is organised as follows: In section~\ref{sec:form} we describe our formalism including the effective model treatment and lay out the notational convention that we follow for the rest of the manuscript. Then in section~\ref{sec:cal} we give the details of our calculation. In section~\ref{sec:res} we discuss the results that we obtain and finally in section~\ref{sec:con} we conclude.

\section{Formalism}
\label{sec:form}
\begin{figure}
\hspace{1cm}
\includegraphics[scale=0.3]{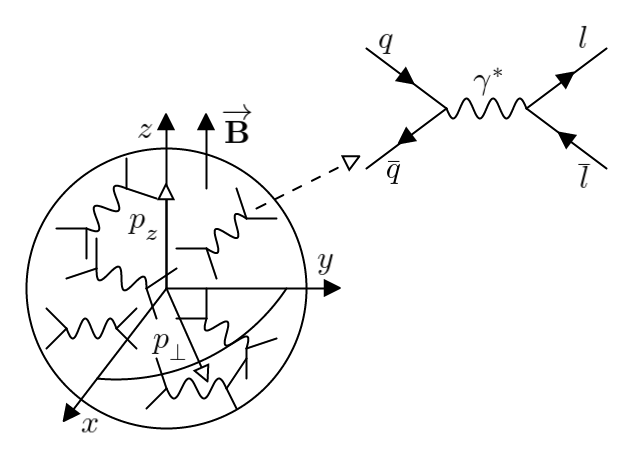}
\caption{A cartoon depicting the dilepton production from a magnetised QCD medium.}
\label{fig:dilepton_cartoon}
\end{figure}
Dileptons are generally produced within the hot and dense fireball in all stages of HIC. Hence it is always more appropriate to evaluate the DR considering arbitrary values of external magnetic field, instead of limiting computations of strong or weak magnetic field approximations~\cite{Bandyopadhyay:2016fyd,Bandyopadhyay:2017raf}. Dileptons are basically lepton pairs generated from the decay of a virtual photon through the annihilation of quark-antiquark pairs.

A schematic picture depicting the generation of lepton pair inside a magnetised QCD medium is shown in Fig.~\ref{fig:dilepton_cartoon}. In our choice of coordinate system, the magnetic field is the $z$ direction, i.e., $\vec {\bf B}\,=\, B \hat{z}$, which is also perpendicular to the reaction plane, $x-y$. The components of any arbitrary four momentum are now separated in two parts \textemdash\, one along the direction of the magnetic field, termed as longitudinal ($\parallel$) component, and the other lying in the $x-y$ plane, termed as transverse or perpendicular ($\perp$) component.

Following this convention, throughout this study we are going to use the following notation for relevant momenta:
\begin{subequations}
\begin{align}
& K^\mu = k_\shortparallel^\mu + k_\perp^\mu;~~ k_\shortparallel^\mu = (k^0,0,0,k^z) ;~~  k_\perp^\mu = (0,k^x,k^y,0),\\
&g^{\mu\nu} = g_\shortparallel^{\mu\nu} + g_\perp^{\mu\nu};~~ g_\shortparallel^{\mu\nu}= \textsf{diag}(1,0,0,-1);~~ g_\perp^{\mu\nu} = \textsf{diag}(0,-1,-1,0),\\
&(K \cdot Q) = (k\cdot q)_\shortparallel - (k\cdot q)_\perp;~~ (k\cdot q)_\shortparallel = k_0q_0-k_zq_z;~~ (k\cdot q)_\perp = (k_xq_x+k_yq_y).
\end{align}
\label{eq:mom_decom}
\end{subequations}
Thus, in our convention, as evident from the above-written relations, $k_z/q_z$ and $k_{\sperp}/q_{\sperp}$ are the spatial parts of the four momentum along and perpendicular to the direction of the magnetic field, respectively. 

\begin{figure}
\includegraphics[scale=0.7]{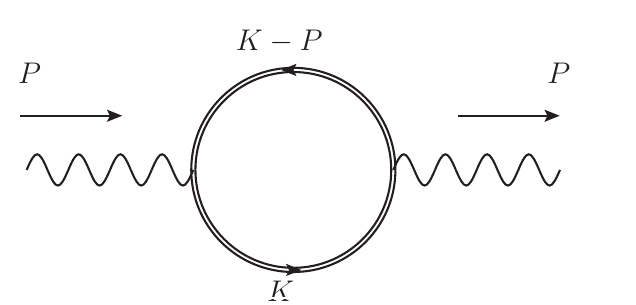}
\caption{Feynman diagram for computing one loop photon self energy or polarization tensor.}
\label{fig:PPT_diag}
\end{figure}

With the knowledge that the lepton pairs are generated from the decay of a virtual photon through the annihilation of quark-antiquark pairs (Fig.~\ref{fig:dilepton_cartoon}), in presence of the external magnetic field it is obvious to assume one of these three pictures where either the quark-antiquark pair, or the lepton pair, or the both pairs can feel the effect of the magnetic field. Since the lepton pairs possess much larger mean free path than the fireball, they come out of it without much interaction as explained in Ref.~\cite{Wang:2022jxx} in detail. Hence, in our current study, we consider the scenario when only the quarks move in a magnetized medium but not the final lepton pairs. In that case, without loss of generality one can neglect the mass of the leptons and the expression of DR can be written as~\cite{Weldon:1990iw},
\begin{align}
\frac{dN}{d^4Xd^4P}\equiv \frac{dR}{d^4P} = \frac{\alpha_{\textsf{EM}}}{12\pi^4}\frac{1}{P^2}\frac{1}{e^{p_0/T}-1}\sum_{f=u,d}\mathsf{Im}\,\Pi_{f\,\,\mu}^{\mu}(P), 
\label{eq:DR}
\end{align}
where $\mathsf{Im}\,\Pi_{f}^{\mu\nu}$ is imaginary part of the photon self energy $\Pi_{f}^{\mu\nu}$ (see Fig.~\ref{fig:PPT_diag}) for flavour $f$ and we will soon go into its detailed description. $\alpha_{\textsf{EM}}$ is the electromagnetic (EM) fine structure constant, $q_f$ is the charge of the fermion of flavor $f$ with $q_u=2/3$ and $q_d=-1/3$, the four-momentum $P\,=\,(p_0,{\bf p})$ which is also related to the invariant mass ($M$) of the lepton pairs as $P^2\,=\,p_0^2-p^2\,=\,M^2$. We have considered the two lightest flavor quarks (up and down; $N_{\sF}=2$) in the present calculation.

The lowest order dilepton rate in the absence of the external magnetic field is called the Born rate which is essential for comparing with our novel results. Considering the case of massless lepton pairs one finds the Born rate~\cite{Cleymans:1986na,Greiner:2010zg} as
\begin{align}
\frac{dN}{d^4xd^4P}\Bigg\vert_{\textsf{Born}} \equiv \frac{dR}{d^4P}\Bigg\vert_{\textsf{Born}} =\frac{5N_c\alpha_{\textsf{EM}}^2}{108\pi^4 p}\frac{T}{\exp( p_0/T)-1}\left(1+\frac{2m^2_{\sF}}{P^2}\right)\log\left[\frac{\left(e^{-(p_0+\mu)/T}+e^{-\omega_{-}/T}\right)\left(e^{-\mu/T}+e^{-\omega_{+}/T}\right)}{\left(e^{- (p_0+\mu)/T}+e^{-\omega_{+}/T}\right)\left(e^{-\mu/T}+e^{-\omega_{-}/T}\right)}\right],
\label{eq:Born_massive_gen}
\end{align} 
where\begin{align}
\omega_{\pm}\equiv \frac{1}{2}\left(p_0\pm p\sqrt{1-\frac{4m_{\sF}^2}{P^2}}\right).
\end{align}

It is appropriate here to digress and describe some of the important terminologies that are associated with explorations of spectral properties like DR. Generally, it is customary to calculate the vector-vector CF and and therefrom estimate different spectral properties including DR as mentioned in the introduction. Apart from that, the study of CF is really fruit-bearing as it also contains information on the mass, width and captures the response of the conserved density fluctuations. Since here, we are solely interested in the calculation of DR which can be directly extracted from the photon self-energy, we skip the detail and just briefly touch on them. The CF can be calculated from the self-energy as $\Pi^{\mu\nu}_{\sF}(P) = q_{\sF}^2\mathcal{C}^{\mu\nu}_{\sF}(P)$ and the very important, spectral function is related to the CF as $\rho_{\sF}^{\sV}(P) = \frac{1}{\pi}\,\mathsf{Im}\,\mathcal{C}_{f\,\,\mu}^{\mu}(P).$ Once they are obtained they can be used to extract many interesting properties of such many particle systems.  Interested readers can find the details in Refs.~\cite{Forster:1975pm,Callen:1951vq,Kubo:1957mj,Weldon:1990iw}

Coming back to the calculation of the photon self-energy in a magnetised medium, we can readily write down the one loop electromagnetic (EM) polarization tensor (for a given flavour $f$) as 
\begin{align}
\Pi^{\mu\nu}_{\sF}(P) = -iN_cq_{\sF}^2\int\!\!\frac{d^4K}{(2\pi)^4}\mathsf{Tr}\left[\gamma^{\mu}S^{\sB}_{\sF}(K)\gamma^{\nu}S^{\sB}_{\sF}(K-P)\right],
\label{eq:photon_pol}
\end{align} 
where we have represented $P$ as the momentum of the external photon line (see Fig.~\ref{fig:PPT_diag}) and $K$ and $K-P(=Q)$ as the fermionic loop momenta. $N_c$ or the number of color is a resultant of a color trace and hence $\mathsf{Tr}$ represents only the Dirac traces.

To calculate the $\Pi^{\mu\nu}$ we use the most general fermionic propagator in a magnetized medium given by Schwinger~\cite{Schwinger:1951nm} and subsequently explored extensively in various previous studies~\cite{Gusynin:1995nb}. As we are interested in evaluating the gauge invariant one loop polarization tensor, the phase factor appearing in Schwinger's original propagator drops out in our calculation and we can directly write down the Schwinger propagator in momentum space as
\begin{align}
S^{\sB}_{\sF}(K) = \exp\left(-\frac{k_{\sperp}^2}{|q_{\sF}B|}\right)\sum_{\ell=0}^{\infty}(-1)^{\ell}\frac{D_{\ell}(K,q_{\sF}B)}{k_{\shp}^2-2\ell|q_{\sF}B|-m_{\sF}^2+i\epsilon}\,,
\label{eq:schwinger_prop_mom}
\end{align}
where 
\begin{align}
D_{\ell}(K,q_{\sF}B)&=(\slashed{k}_{\shp}+m_{\sF})\left\{L_{\ell}\left(\frac{2k_{\sperp}^2}{|q_{\sF}B|}\right)\left[\mathbbm{1}-i\gamma^1\gamma^2\textsf{sgn}(q_{\sF}B)\right]-L_{\ell-1}\left(\frac{2k_{\sperp}^2}{|q_{\sF}B|}\right)\left[\mathbbm{1}+i\gamma^1\gamma^2\textsf{sgn}(q_{\sF}B)\right]\right\}\nn
&+4\slashed{k}_{\sperp}L^{1}_{\ell-1}\left(\frac{2k_{\sperp}^2}{|q_{\sF}B|}\right).
\end{align}
Here, $m_f$ and $q_f$ are the mass and charge of the fermion of flavor $f$ respectively. $\ell$ denotes the Landau level index and $L_{\ell}^{\alpha} (x)$ is the generalized Laguerre polynomial written as
\begin{align}
(1-z)^{-(\alpha+1)}\exp\left(\frac{xz}{z-1}\right) = \sum_{\ell =0}^{\infty} L_{\ell}^\alpha(x) z^{\ell}. \label{eq:Laguerre_generating}
\end{align}
We are now equipped with all the necessary machinery to calculate the photon self-energy and therefrom the rate in presence of an arbitrary magnetic field. Before going into the details of that, we dedicate the next subsection to build up the necessary formalism for the effective model that we use. 

\subsection{In the ambit of effective model}
\label{ssec:form_eff}
To investigate the effect of phase transition (PT), particularly the chiral one, on the rate we utilise an effective model and revisit the calculation. As an quintessential example, we consider the Nambu\textemdash Jona-Lasinio model~\cite{Nambu:1961tp,Nambu:1961fr}. In such effective model, the chiral symmetry is spontaneously broken by the non-zero chiral condensate at low temperature and gets restored at a temperature higher than the PT temperature ($T_{CO}$). In turn, this chiral condensate lends mass to the quarks, which then attain effective masses and behave as quasi-quarks. As the masses of the quarks directly go as the inputs for the rate calculation, the rate is bound to be affected by the PT.

The effect of chiral PT or so to speak the quasi-quarks on the rate has been previously examined in Ref.~\cite{Islam:2014sea} in absence of magnetic field. Similar studies in presence of $eB$ was first executed in Ref.~\cite{Islam:2018sog} and later extended in Refs.~\cite{Ghosh:2020xwp,Chaudhuri:2021skc}. One important aspect of the PT in presence of magnetic field is the IMC effect~\cite{Bali:2012zg,Bali:2011qj}, which is missing in all these effective model implementations. We try to incorporate the IMC effect on the rate by using a version of the model developed in~\cite{Farias:2016gmy}.

The consideration of IMC effect is particularly of greater importance in the region of transition temperature, where the increase of $eB$ will decrease the strength of the condensate as opposed to its increase at lower or higher temperature values~\cite{Bali:2012zg,Bali:2011qj}. Thus the DR will be affected by the IMC effect through the quasi-quarks, depending on the temperature of evaluation.

The Lagrangian density for the isospin-symmetric two-flavor NJL model in presence of an external electromagnetic field ($A^\mu$) is given by
\be
\mathcal{L}_{\rm NJL}= -\frac{1}{4} F^{\mu\nu}F_{\mu\nu}
+ \bar{\psi}\left(i\slashed{D}-m\right)\psi
+ G_S\left[ (\bar{\psi}\psi)^2+(\bar{\psi}i\gamma_5{\vec\tau}\psi)^2\right],
\label{NJL_lag}
\ee
where $m = {\rm diag}(m_f)$ is the quark-mass with $m_u=m_d$ for isospin-symmetric matter,
$D_\mu = \partial_\mu - iq A_\mu$ is the covariant derivative, $F_{\mu\nu} = \partial_\mu A_\nu - \partial_\nu A_\mu$, and $\vec \tau = (\tau^1, \tau^2, \tau^3)$ are the isospin Pauli matrices. $q = {\rm diag}(q_f) ={\rm diag}(q_u = 2e/3, q_d =-e/3)$ is the charge matrix in the flavor space and $G_S$ is the scalar coupling constant of the NJL model. 

It is important to note that NJL, being a non-renormalisable model, needs to be regularised. We use a three momentum cut-off $(\Lambda)$ here for that purpose. The coupling constant $G_S$, the current quark mass $m$ and $\Lambda$ are the parameters of the model which are fitted by using zero temperature QCD observables: pion mass $(m_\pi)$, pion decay constant $(f_\pi)$ and the condensate $(\sigma)$. The details of the fitting and corresponding values can be found in Ref.~\cite{Farias:2016gmy}.

Within the mean field approximation, the gap equation for the constituent quark mass $\mathcal{M}={\rm diag}(\mathcal{M}_f)$ at finite temperature $T$ and in the presence of a magnetic field~$B$ is given by
\begin{align}
\mathcal{M} = m - 2 G_S  \sum_{f=u,d}\langle \bar{\psi}_f\psi_f\rangle,
\label{Eq:Gap_B}
\end{align}
where $\langle \bar{\psi}_f\psi_f\rangle$ is the quark condensate of flavor~$f$. Eq.~\ref{Eq:Gap_B} depicts how the quasi-quark attains the effective mass $(\mathcal{M})$ depending on the values of the quark condensate and the scalar coupling constant, $G_S$.

As mentioned earlier, in the present work, we plan to include the effects of the IMC phenomenon on the quasi particle effective/constituent mass $\mathcal{M}$. In the standard NJL model, where we find only MC throughout the temperature range, $G_S$ does not depend on temperature or magnetic field and remains a constant. Here we attain the desired IMC effect by using a medium dependent scalar coupling $G_S(eB,T)$~\cite{Farias:2016gmy}:
\be
G_S(eB,T) = c(eB)\left[1-\frac{1}{1+e^{\beta(eB)[T_a(eB)-T]}} \right]+s(eB),
\label{GBT}
\ee
where $c(eB)$, $\beta(eB)$, $T_a(eB)$ and $s(eB)$ depend only on the magnitude of $B$ and their values for selected strengths of $eB$ are given in Table~1 of Ref.~\cite{Farias:2016gmy}. All following numerical results in the ambit of effective model refer to this parametrization. 

In the following section we will provide details of our calculation, starting from the one loop EM polarization tensor until the evaluation of the total DR.


\section{calculation}
\label{sec:cal}

We start the present section with the expression of photon self-energy, where we have contracted the indices $\mu$ and $\nu$ in Eq.~\eqref{eq:photon_pol}, resulting in 
\begin{align}
\Pi^\mu_{\mu,\sF}(P) = -iN_cq_{\sF}^2\int\!\!\frac{d^4K}{(2\pi)^4}\mathsf{Tr}\left[\gamma^{\mu}S^{\sB}_{\sF}(K)\gamma_{\mu}S^{\sB}_{\sF}(K-P)\right].
\label{eq:photon_pol_contr}
\end{align} 
One can now perform the trace in Eq.~\eqref{eq:photon_pol_contr}, which is given as ,
\begin{align}
\textsf{Tr}\left[\gamma^{\mu}D_{\ell}(K,q_{\sF}B)\gamma_{\mu}D_{\ell}(K-P,q_{\sF}B)\right] &= 16\left[(k.q)_{\shp}-m_{\sF}\right]
\left[L_{\ell}\left(\frac{2k_{\sperp}^2}{|q_{\sF}B|}\right)L_{n-1}\left(\frac{2q_{\sperp}^2}{|q_{\sF}B|}\right) + L_{\ell-1}\left(\frac{2k_{\sperp}^2}{|q_{\sF}B|}\right)L_{n}\left(\frac{2q_{\sperp}^2}{|q_{\sF}B|}\right)\right]\nn
&+16m_{\sF}^2\left[L_{\ell}\left(\frac{2k_{\sperp}^2}{|q_{\sF}B|}\right)L_{n}\left(\frac{2q_{\sperp}^2}{|q_{\sF}B|}\right) + L_{\ell-1}\left(\frac{2k_{\sperp}^2}{|q_{\sF}B|}\right)L_{n-1}\left(\frac{2q_{\sperp}^2}{|q_{\sF}B|}\right)\right] \nn
&+128\,(k.q)_{\sperp}L_{\ell-1}^1\left(\frac{2k_{\sperp}^2}{|q_{\sF}B|}\right)L_{n-1}^1\left(\frac{2q_{\sperp}^2}{|q_{\sF}B|}\right).
\end{align}
The quantity, $(k.q)_{\shp}-m_{\sF}^2 =\frac{1}{2}\left[k^2_{\shp}+(k-p)_{\shp}^2-p^2_{\shp}-2m_{\sF}^2\right] =\Delta_{f,\ell,k}^{\shp}+\Delta_{f,n,q}^{\shp}+(\ell+n)|q_{\sF}B|-\frac{1}{2}p_{\shp}^2$,
where we define $\Delta_{f,\ell,k}^{\shp} = k_{\shp}^2-m_{f,\ell}^2$, $\Delta_{f,n,q}^{\shp} = q_{\shp}^2-m_{f,n}^2$, $m_{f,\ell}=\sqrt{2\ell |q_{\sF}B|+m_{\sF}^2}$ and $m_{f,n}=\sqrt{2n|q_{\sF}B|+m_{\sF}^2}$. One also needs to be mindful here while calculating the integral. Because of the decomposition of momentum into parallel and perpendicular components as given in Eqs.~\ref{eq:mom_decom}, the integral over the four momentum are also separated as,
\begin{align}
\int\frac{d^4K}{(2\pi)^4} = \int\frac{d^2k_\shortparallel}{(2\pi)^2}\times \int\frac{d^2k_\perp}{(2\pi)^2}.
\end{align}

Putting everything together in Eq.~\eqref{eq:photon_pol_contr} we get,
\begin{align}
\Pi^{\mu}_{\mu,f}(P) &= -i16N_cq_{\sF}^2\left[\sum_{\ell,n = 0}^{\infty}(-1)^{\ell+n}\mathcal{N}_{f,\ell,n}(p_{\shp}^2,p_{\sperp}^2)\int\!\!\frac{d^2k_{\shp}}{(2\pi)^2}\frac{1}{\Delta_{f,\ell,k}^{\shp}\Delta_{f,n,q}^{\shp}} \right. \nn
&\left.+\sum_{\ell,n=0}^{\infty}(-1)^{\ell+n}\left[\mathcal{J}^{(0)}_{f,\ell,n-1}(p_{\sperp}^2)+\mathcal{J}^{(0)}_{f,\ell-1,n}(p_{\sperp}^2)\right]\int\!\!\frac{d^2k_{\shp}}{(2\pi)^2}\left(\frac{1}{\Delta_{f,\ell,k}^{\shp}}+\frac{1}{\Delta_{f,n,q}^{\shp}}\right)\right].  \label{eq:Pimumu}
\end{align}
For the compactification of the expression in the above equation we have defined,
\begin{align}
\mathcal{N}_{f,\ell,n}(p_{\shp}^2,p_{\sperp}^2) &\equiv \left[(\ell+n)|q_{\sF}B|-\frac{1}{2}p_{\shp}^2\right]\left[\mathcal{J}^{(0)}_{f,\ell-1,n}(p_{\sperp}^2) + \mathcal{J}^{(0)}_{f,\ell,n-1}(p_{\sperp}^2)\right]+m_{\sF}^2\left[\mathcal{J}^{(0)}_{f,\ell,n}(p_{\sperp}^2)
+\mathcal{J}^{(0)}_{f,\ell-1,n-1}(p_{\sperp}^2)\right]\nn
&+8\mathcal{J}_{f,\ell-1,n-1}^{(1)}(p_{\sperp}^2),
\end{align}
where 
\begin{align}
\mathcal{J}^{(\alpha)}_{f,\ell,n}(p_{\sperp}^2) \equiv \int\!\!\frac{d^2k_{\sperp}}{(2\pi)^2}\exp\left(-\frac{k_{\sperp}^2+q_{\sperp}^2}{|q_{\sF}B|}\right)\,(k.q)_{\sperp}^{\alpha}\,L_{\ell}^{\alpha}\left(\frac{2k_{\sperp}^2}{|q_{\sF}B|}\right)L_{n}^{\alpha}\left(\frac{2q_{\sperp}^2}{|q_{\sF}B|}\right). \label{eq:Jalpha}
\end{align}
Note that, we need to find $\textsf{Im}\Pi^{\mu}_{\mu,f}(p_0+i\epsilon,\bm{p})$, with $\epsilon\rightarrow 0^{+}$. The last term in the square bracket of R.H.S of Eq.~\eqref{eq:Pimumu} can be dropped as its $p_0$ dependence can be removed by changing variable from $k_0\rightarrow k_0^{\prime}=p_0-k_0$ and $k_z\rightarrow k_z^{\prime}=p_z-k_z$.
Now,  we are interested to perform the integration over the perpendicular momenta shown in Eq.~\eqref{eq:Jalpha} only for $\alpha =0,1$. It can be performed analytically as shown in Appendix~\ref{app:perp_intg}. The results for $\alpha =0, 1$ are quoted below:
\begin{align}
\mathcal{J}^{(0)}_{f,\ell,n}(p_{\sperp}^2) = 
\begin{cases}
(-1)^{\ell+n}\dfrac{|q_{\sF}B|}{8\pi}\dfrac{\textsf{min}(\ell,n)!}{\textsf{max}(\ell,n)!}\xi^{|\ell-n|}e^{-\xi}\left[L_{\textsf{min}(\ell,n)}^{|\ell-n|}(\xi)\right]^2 & \text{for  } \,\,\,p_{\sperp}\neq 0\\
\dfrac{|q_{\sF}B|}{8\pi}\delta_{\ell, n} & \text{for  }\,\,\, p_{\sperp} = 0  
\end{cases} \label{eq:J0}
\end{align}
and
\begin{align}
\mathcal{J}^{(1)}_{f,\ell,n}(p_{\sperp}^2) =
\begin{cases}
 (-1)^{\ell+n}\dfrac{|q_{\sF}B|^2}{16\pi}\dfrac{\big(\textsf{min}(\ell,n)+1\big)!}{\textsf{max}(\ell,n)!}\xi^{|\ell-n|}e^{-\xi}L_{\textsf{min}(\ell,n)}^{|\ell-n|}(\xi)L_{\textsf{min}(\ell,n)+1}^{|\ell-n|}(\xi) & \text{for  } \,\,\,p_{\sperp}\neq 0\\
\dfrac{|q_{\sF}B|^2}{16\pi}(\ell+1)\delta_{\ell+1, n+1} & \text{for  } \,\,\,p_{\sperp} = 0\,, 
 \end{cases} \label{eq:J1}
\end{align}
respectively, where $\xi = \dfrac{p_{\sperp}^2}{2|q_{\sF}B|}$.  It is easy to check the result for $p_{\sperp}=0$ and from $p_{\sperp}\neq 0$ in Eq.~\eqref{eq:J0} and Eq.~\eqref{eq:J1}. For $\ell\neq n$ and $p_{\sperp}=0$,  both the perpendicular integral vanishes.  When $\ell$ and $n$ take the same integer value, we must first put $\ell=n$ and then take the limit $p_{\sperp}\rightarrow 0$ or $\xi\rightarrow 0$ to obtain the desired result. To arrive at this desired limit we use the property of Laguerre polynomial $L^0_{i}(\xi\rightarrow 0)=1$, for $i=0,1,2,\cdots$.  
 
In the imaginary time formalism of thermal field theory, the temperature effect is taken into account by replacing the $k_0$ integral by discrete frequency sum as follows: 
\begin{align}
\int\limits_{-\infty}^{\infty}\frac{dk_0}{2\pi} \longrightarrow iT\sum_{n=-\infty}^{\infty}\qquad{\rm with}\qquad k_0 = i(2n+1)\pi T + \mu.
\end{align}
Then we analytically continue the result to real $p_0$.
After this,  from the first term of Eq.~\eqref{eq:Pimumu}, we get
\begin{align}
\mathsf{Im}\Pi^{\mu}_{f\,\,\mu}(P) &=-4N_cq_{\sF}^2\sum_{\ell,n=0}^{\infty}(-1)^{\ell+n}\nn &\times\mathcal{N}_{f,\ell,n}(p_{\shp}^2,p_{\sperp}^2)\int\limits_{-\infty}^{\infty}\!\!\frac{dk_z}{2\pi}\sum_{s_1=\pm 1}\sum_{s_2=\pm 1}\frac{s_1s_2}{E_{f,\ell,k}E_{f,n,q}}\mathsf{Im}\left(\frac{1-\wt{n}_{+}(s_1E_{f,\ell,k})-\wt{n}_{-}(s_2E_{f,n,q})}{p_0-s_1E_{f,\ell,k}-s_2E_{f,n,q}+i\epsilon}\right)
\end{align}
Using the identity 
\begin{align}
\textsf{Im}\frac{1}{x+i\epsilon} = -\pi\delta(x),
\end{align}
we get
\begin{align}
\rho^{\sV}_{\sF}(P) &= 4N_c\sum_{\ell,n=0}^{\infty}(-1)^{\ell+n}\mathcal{N}_{f,\ell,n}(p_{\shp}^2,p_{\sperp}^2)\nn
&\times\sum_{s_1=\pm 1}\sum_{s_2=\pm 1}\int\limits_{-\infty}^{\infty}\!\!\frac{dk_z}{2\pi}\frac{1-\wt{n}_{+}(s_1E_{f,\ell,k})-\wt{n}_{-}(s_2E_{f,n,q})}{s_1s_2E_{f,\ell,k}E_{f,n,q}}\delta(p_0-s_1E_{f,\ell,k}-s_2E_{f,n,q}).\label{eq:expression_tot_rhoV}
\end{align}
Here we define $E_{f,\ell,k} \equiv \sqrt{k_z^2+m_{f,\ell}^2}$\,, $E_{f,n,q} \equiv \sqrt{(k_z-p_z)^2+m_{f,n}^2}$ and
\begin{align}
\wt{n}_{\pm}(\mathcal{E}) = \frac{1}{\exp(\frac{\mathcal{E}\mp\mu}{T})+1}
\end{align}
is the Fermi-Dirac distribution function.
There are four delta functions associated with the dilepton production rate (DR), $\delta \left(p_0-s_1E_{f,\ell,k}-s_2E_{f,n,q}\right)$.
The energy component $p_0$ is taken to be positive to consider the emission of virtual photon $\gamma^{*}$ as negative $p_0$ should be associated with the absorption of $\gamma^{*}$. Thus, we shall restrict ourselves to the $p_0>0$ case only . Now, $s_1=s_2=1$, $s_1=-s_2=1$ and $s_1=-s_2=-1$ corresponds to annihilation process ($q+\bar{q}\rightarrow \gamma^{*}$), particle decay process ($q\rightarrow q+\gamma^{*}$) and anti-particle decay process ($\bar{q}\rightarrow \bar{q}+\gamma^{*}$), respectively.
There will be no contribution from $s_1=s_2=-1$ case as the argument will always be greater than zero in this case owing to the fact that $p_0$, $E_{f,\ell,k}$ and $E_{f,n,q}$ are always positive definite. So we will not consider that scenario in the rest of the article.\\ \\
Now, we solve 
\begin{align}
F_{f,s_1,s_2}(k_z)\equiv p_0 -s_1E_{f,\ell,k}-s_2E_{f,n,q} = 0 \label{eq:energy_eq_gen}
\end{align}
for $k_z$ to perform the $k_z$ integration. For real $k_z$, the four delta functions are satisfied in different kinematic regimes. The solution of $k_z$ in Eq.~\eqref{eq:energy_eq_gen} is 
\begin{align}
k_{z}^{\sigma} = \frac{1}{2p_{\shp}^2}\left[p_z\left(p_{\shp}^2+2 (\ell-n)|q_fB|\right)+\sigma p_0\lambda^{1/2}(p_{\shp}^2,m_{f,\ell}^2,m_{f,n}^2)\right], \label{eq:kz_sol_gen}
\end{align}
where $\lambda(a,b,c)=a^2+b^2+c^2-2ab-2bc-2ca$ is the K{\" a}ll{\' e}n function and $\sigma=\pm 1$. To obtain $E_{f,\ell,k^{\sigma}}\equiv E^{\sigma}_{f,\ell,k}$ we take $s_2E_{f,n,q}$ to the R.H.S of Eq.~\eqref{eq:energy_eq_gen} and square both sides and rewrite it as
\begin{align}
E^{\sigma}_{f,\ell,k} = \frac{1}{2p_0s_1}\left[p_0^2+(k_z^{\sigma})^2+m_{f,\ell}^2-(k_z^{\sigma}-p_z)^2-m_{f,n}^2\right].
\end{align}   Substituting $k_z^{\sigma}$ from Eq.~\eqref{eq:kz_sol_gen} in the above equation and after simplifying we get,
\begin{align}
E^{\sigma}_{f,\ell,k} = s_1\mathcal{E}^{\sigma}_{f,\ell,k} = s_1\frac{p_0\left(p_{\shp}^2+2 (\ell-n)|q_fB|\right)+\sigma |p_z|\lambda^{1/2}(p_{\shp}^2,m_{f,\ell}^2,m_{f,n}^2)}{2p_{\shp}^2}.
\end{align}
Similarly, to obtain $E^{\sigma}_{f,n,q}$, we take $s_1E_{f,\ell,k}$ to the R.H.S of Eq.~\eqref{eq:energy_eq_gen} and follow a similar procedure to obtain
\begin{align}
E^{\sigma}_{f,n,q} = s_2\mathcal{E}^{\sigma}_{f,n,q} = s_2\frac{p_0\left(p_{\shp}^2+2 (n-\ell)|q_fB|\right)-\sigma |p_z|\lambda^{1/2}(p_{\shp}^2,m_{f,\ell}^2,m_{f,n}^2)}{2p_{\shp}^2}.
\end{align}  
We make use of the identity to simplify delta function
\begin{align}
\int dx\,\delta(f(x)) = \sum_{i}\frac{\delta(x-x_i)}{|f^{\prime}(x_i)|},\qquad \text{where}\; x_i\,\,\text{are the roots of }\,f(x)=0. \label{eq:Fs1s2}
\end{align}
Thus we have,
\begin{align}
\Bigg\vert \frac{\partial}{\partial k_z}F_{s_1,s_2}(k_z)\Bigg\vert_{k_z=k^{\sigma}_z} = \frac{\lambda^{1/2}\left(p_{\shp}^2,m^2_{f,\ell},m^2_{f,n}\right)}{2E^{\sigma}_{f,\ell,k}E^{\sigma}_{f,n,q}}.
\end{align}
Now, following Eq.~\eqref{eq:Fs1s2}, we have
\begin{align}
&\delta(p_0-s_1E_{f,\ell,k}-s_2E_{f,n,q})  \nn
&=\begin{cases}
\sum_{\sigma} \dfrac{2E^{\sigma}_{f,\ell,k}E^{\sigma}_{f,n,q}}{\lambda^{1/2}\left(p_{\shp}^2,m^2_{f,\ell},m^2_{f,n}\right)}\delta(k_z-k_z^{\sigma})\Theta\left(p_{\shp}^2-(m_{f,\ell}+m_{f,n})^2\right) & \text{for  }s_1=s_2=1
\\
\sum_{\sigma} \dfrac{2E^{\sigma}_{f,\ell,k}E^{\sigma}_{f,n,q}}{\lambda^{1/2}\left(p_{\shp}^2,m^2_{f,\ell},m^2_{f,n}\right)}\delta(k_z-k_z^{\sigma})\Theta\left((m_{f,\ell}-m_{f,n})^2-p_{\shp}^2\right) &  \text{for  }s_1=-s_2=1 \text{ with } \ell > n \\
\sum_{\sigma} \dfrac{2E^{\sigma}_{f,\ell,k}E^{\sigma}_{f,n,q}}{\lambda^{1/2}\left(p_{\shp}^2,m^2_{f,\ell},m^2_{f,n}\right)}\delta(k_z-k_z^{\sigma})\Theta\left((m_{f,\ell}-m_{f,n})^2-p_{\shp}^2\right) &  \text{for  }s_1=-s_2=-1 \text{ with } \ell < n\,. \\
\end{cases} \label{eq:dirac_delta_simplified}
\end{align}
Note that, for $s_1,s_2\in \{-1,1\}$,  it is actually four equations that are being represented by Eq.~\eqref{eq:energy_eq_gen} in a compact way.  For given values of each parameters, i.e. $p_0$, $p_z$, $\ell$, $n$, $|q_{\sF}B|$ and $m_{\sF}$, one can get real solutions of $k_z$ just using one equation represented by Eq.~\eqref{eq:energy_eq_gen}.
So for each case, the theta functions along with $\ell\gtrless n$ condition decide which equation gives real $k_z$ as solution.  The conditions represented by theta function are discussed in details in App.~\eqref{app:unitary_cuts} and App.\eqref{app:landau_cut}. One must maintain the condition $p_0^2-p_{\sperp}^2-p_z^2>0$ to preserve the causal  behaviour of the theory. Now we have $p_0-s_1E_{\ell,k}^{\sigma}-s_2E_{n,q}^{\sigma}=0$, from where $p_0>0$ gives $s_1E_{\ell,k}^{\sigma}-s_2E_{n,q}^{\sigma}>0$. For $s_1=-s_2=1$,  it gives $\ell - n >\frac{|p_z|\lambda^{1/2}\left(p_{\shp}^2, m^2_{f,\ell},m^2_{f,n}\right)}{2|q_{\sF}B|p_0}$. Also for $s_1=-s_2=-1$, it gives $\ell - n <-\frac{|p_z|\lambda^{1/2}\left(p_{\shp}^2, m^2_{f,\ell},m^2_{f,n}\right)}{2|q_{\sF}B|p_0}$.  Thus it is evident that for $\ell>n$ the quark decay processes is being satisfied and for $\ell<n$ the antiquark decay process is being satisfied. 
Finally using Eq.~\eqref{eq:DR},  Eq.~\eqref{eq:expression_tot_rhoV} and Eq.~\eqref{eq:dirac_delta_simplified}, below we write down the expressions of the dilepton rates for three different processes, respectively.
\subsection{$q+\bar{q}\rightarrow \gamma^*$ process}

The rate for the quark-antiquark annihilation process is given as,
\begin{align}
& \frac{dN}{d^4xd^4P}\Bigg\vert_{q+\bar{q}\rightarrow \gamma^*} =\frac{\alpha_{\textsf{EM}}^2}{3\pi^3}\frac{1}{p_0^2-p^2}\frac{1}{\exp\left(\frac{p_0}{T}\right)-1}\sum_{\sF=u,d}\left(\frac{q_{f}}{e}\right)^24N_c\sum_{\ell=0}^{\infty}\sum_{n=0}^{\infty}(-1)^{\ell+n}\mathcal{N}_{f,\ell,n}\left(p^2_{\shp},p_{\sperp}^2\right)\frac{\Theta\left(p_{\shp}^2-\left(m_{f,\ell}+m_{f,n}\right)^2\right)}{\lambda^{1/2}\left(p_{\shp}^2,m^2_{f,\ell},m^2_{f,n}\right)}\nn
&\hspace*{5cm}\times\left[2-\tilde{n}_{+}\left(\mathcal{E}^{+}_{f,\ell,k}\right)
-\tilde{n}_{-}\left(\mathcal{E}^{+}_{f,n,q}\right)-\tilde{n}_{+}\left(\mathcal{E}^{-}_{f,\ell,k}\right)
-\tilde{n}_{-}\left(\mathcal{E}^{-}_{f,n,q}\right)\right].
\end{align}
From the positivity condition of the argument of theta function, we have $\sqrt{p_{\shp}^2}\geq m_{f,\ell}+m_{f,n}$.  After simplifying, it gives the upper limit on $n$ as follows
\begin{align}
n\leq \ell +\left\lfloor\frac{p_{\shp}^2-2m_{f,\ell}\sqrt{p_{\shp}^2}}{2|q_{\sF}B|}\right\rfloor .
\end{align}
Now the condition $n\geq 0$ gives the upper limit of maximum value of $\ell$ as
\begin{align}
\ell \leq \left\lfloor\frac{p_{\shp}^2-2m_{\sF}\sqrt{p_{\shp}^2}}{2|q_{\sF}B|}\right\rfloor .
\end{align}
In the last two above equations, the bracketed quantities indicate their respective floors.
\subsection{$q\rightarrow q+\gamma^*$ process}

The rate for the quark decay process is given as, 
\begin{align}
& \frac{dN}{d^4xd^4P}\Bigg\vert_{q\rightarrow q+\gamma^*} =\frac{\alpha_{\textsf{EM}}^2}{3\pi^3}\frac{1}{p_0^2-p^2}\frac{1}{\exp\left(\frac{p_0}{T}\right)-1}\sum_{\sF=u,d}\left(\frac{q_{f}}{e}\right)^24N_c\sideset{}{'}\sum_{\ell,n=0}^{\infty}(-1)^{\ell+n}\mathcal{N}_{f,\ell,n}\left(p^2_{\shp},p_{\sperp}^2\right)\frac{\Theta\left(\left(m_{f,\ell}-m_{f,n}\right)^2-p_{\shp}^2\right)}{\lambda^{1/2}\left(p_{\shp}^2,m^2_{f,\ell},m^2_{f,n}\right)}\nn
&\hspace{5cm}\times\left[\tilde{n}_{+}\left(\mathcal{E}^{+}_{f,\ell,k}\right)-\tilde{n}_{+}\left(-\mathcal{E}^{+}_{f,n,q}\right)+\tilde{n}_{+}\left(\mathcal{E}^{-}_{f,\ell,k}\right)-\tilde{n}_{+}\left(-\mathcal{E}^{-}_{f,n,q}\right)\right].
\end{align}

In the sum over the LL in the above equation,  the prime in the summation sign denotes the condition $\ell >n$. In this case,  the condition $\left(m_{f,\ell}-m_{f,n}\right)^2-p_{\shp}^2 \geq 0$ and $\ell > n$ lead to the upper limit of 
\begin{align}
n\leq \ell +\left\lfloor\frac{p_{\shp}^2-2m_{f,\ell}\sqrt{p_{\shp}^2}}{2|q_{\sF}B|}\right\rfloor 
\end{align}
and lower limit of $\ell$ is
\begin{align}
\ell \geq \left\lfloor\frac{p_{\shp}^2+2m_{\sF}\sqrt{p_{\shp}^2}}{2|q_{\sF}B|}\right\rfloor .
\end{align}
\subsection{$\bar{q}\rightarrow \bar{q}+\gamma^*$ process}

The rate for the antiquark decay process is written as,
\begin{align}
& \frac{dN}{d^4xd^4P}\Bigg\vert_{\bar{q}\rightarrow \bar{q}+\gamma^*} =\frac{\alpha_{\textsf{EM}}^2}{3\pi^3}\frac{1}{p_0^2-p^2}\frac{1}{\exp\left(\frac{p_0}{T}\right)-1}\sum_{\sF=u,d}\left(\frac{q_{f}}{e}\right)^24N_c\sideset{}{'}\sum_{\ell,n=0}^{\infty}(-1)^{\ell+n}\mathcal{N}_{f,\ell,n}\left(p^2_{\shp},p_{\sperp}^2\right)\frac{\Theta\left(\left(m_{f,\ell}-m_{f,n}\right)^2-p_{\shp}^2\right)}{\lambda^{1/2}\left(p_{\shp}^2,m^2_{f,\ell},m^2_{f,n}\right)}\nn
&\hspace{5cm}\times\left[\tilde{n}_{-}\left(\mathcal{E}^{+}_{f,n,q}\right)-\tilde{n}_{-}\left(-\mathcal{E}^{+}_{f,\ell,k}\right)+\tilde{n}_{-}\left(\mathcal{E}^{-}_{f,n,q}\right)-\tilde{n}_{-}\left(-\mathcal{E}^{-}_{f,\ell,k}\right)\right].
\end{align}

In this case the condition $n>\ell$ must be satisfied and which is denoted by the prime on the sum. Similarly, for the anti-particle decay processes the upper limit of $\ell$ is 
\begin{align}
\ell\leq n + \left\lfloor\frac{p_{\shp}^2-2m_{f,n}\sqrt{p_{\shp}^2}}{2|q_{\sF}B|}\right\rfloor 
\end{align}
and the lower limit of $n$ is
\begin{align}
n \geq \left\lfloor\frac{p_{\shp}^2+2m_{\sF}\sqrt{p_{\shp}^2}}{2|q_{\sF}B|}\right\rfloor.
\end{align}

Hence the total DR becomes 
\begin{align}
& \frac{dN}{d^4xd^4P}\Bigg\vert_{\sf{Total}} =\frac{dN}{d^4xd^4P}\Bigg\vert_{q+\bar{q}\rightarrow \gamma^*} + \frac{dN}{d^4xd^4P}\Bigg\vert_{q\rightarrow q+\gamma^*} +\frac{dN}{d^4xd^4P}\Bigg\vert_{\bar{q}\rightarrow \bar{q}+\gamma^*}.
\label{eq:DR_total}
\end{align}

In the next section we discuss various physical consequences of Eq.~\eqref{eq:DR_total}.


\section{Results}
\label{sec:res}
\begin{figure}
\begin{center}
\includegraphics[scale=0.4]{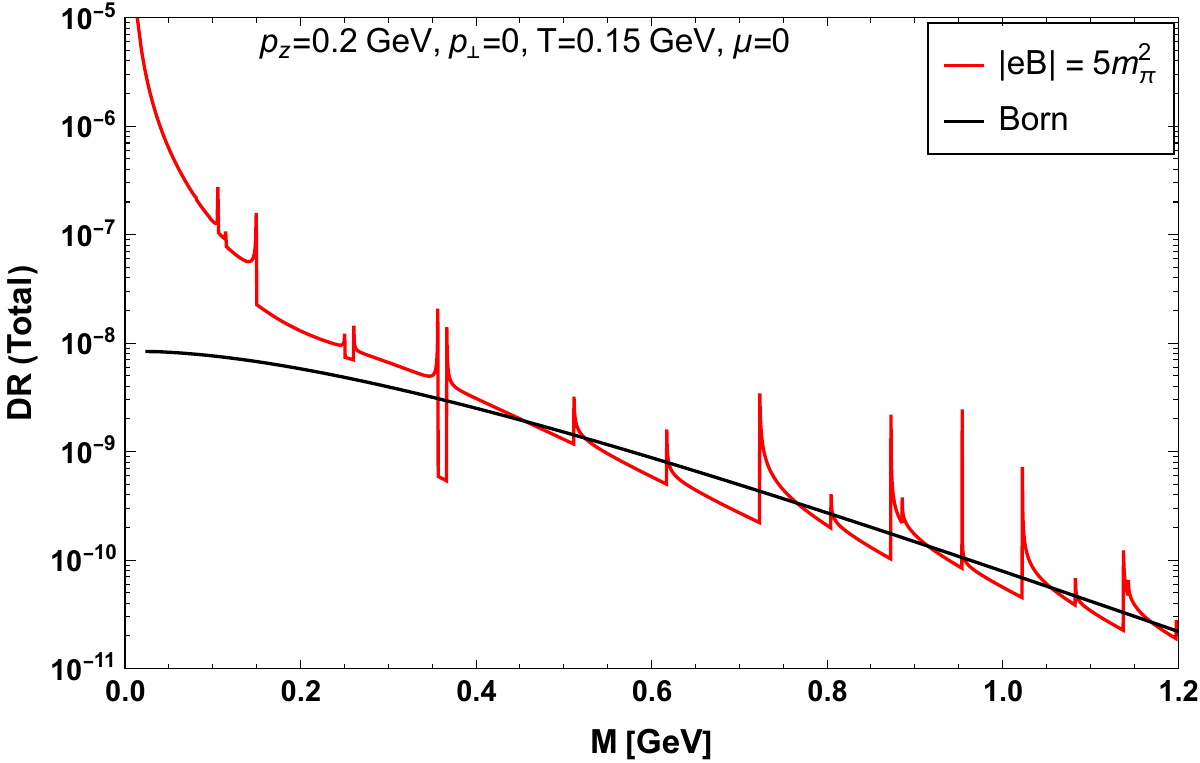}
\includegraphics[scale=0.4]{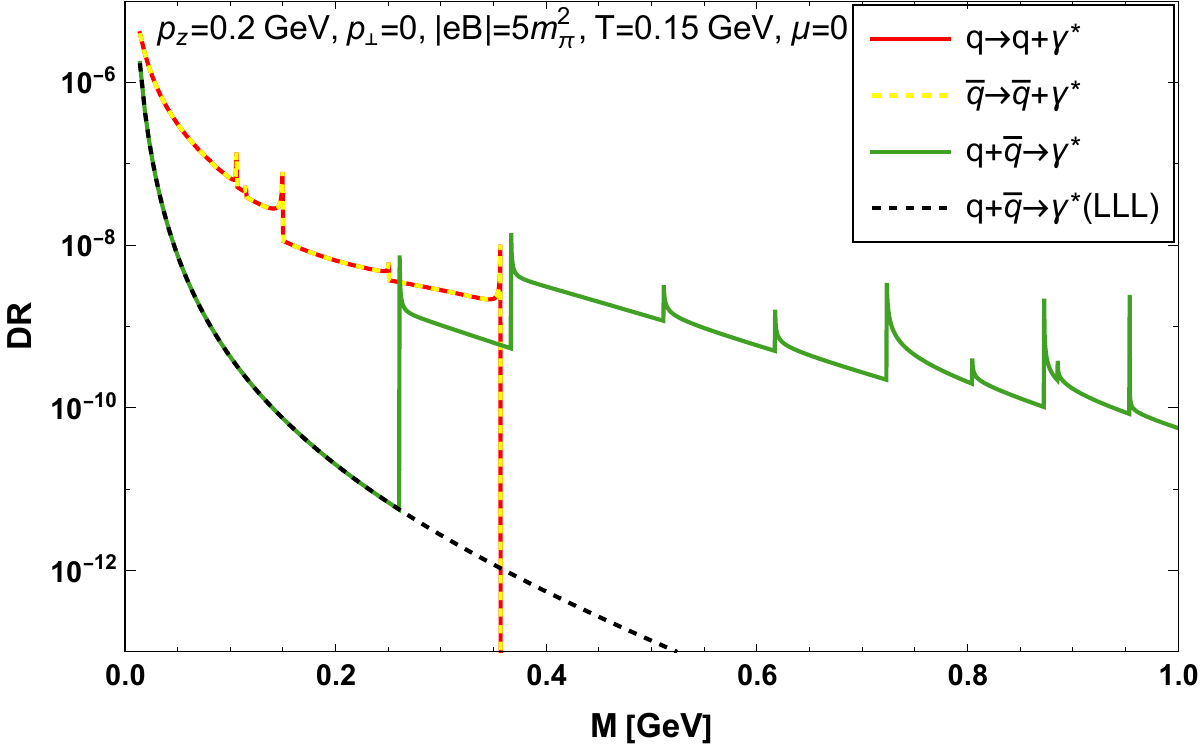}
\caption{(Color online) In the left panel we have the plot of DR as a function of invariant mass for $eB=5\,m_\pi^2$ with $p_T$ being zero. In the right panel the contribution coming from different processes are shown separately along with the LLL approximated rate in the black dashed line.}
\label{fig:dr_diff_process}
\end{center}
\end{figure}
After evaluating the expression for the total DR in the previous section, in this section we focus on exploring different physical aspects of the same along with their implications. To begin with, we show an important plot of DR in Fig~\ref{fig:dr_diff_process} as a function of invariant mass ($M$) for a given nonzero value of magnetic field ($eB=5m_\pi^2$) and compare our result with the Born rate $(eB=0)$ obtained using Eq.~\eqref{eq:Born_massive_gen} while keeping the values of the other relevant parameters same for both. For this plot we keep the transverse momentum ($p_{\sperp}$) equal to zero. Plots with non-zero values of $p_{\sperp}$ are important generalisation of our result and will be discussed in details later in this section. The other parameter values used in this plot are: the momentum along the longitudinal direction of $eB$ $(p_z)$ $=0.2$ GeV, temperature $(T)$ $=0.15$ GeV and chemical potential ($\mu$) $=0$.
\begin{figure}
\begin{center}
\includegraphics[scale=0.4]{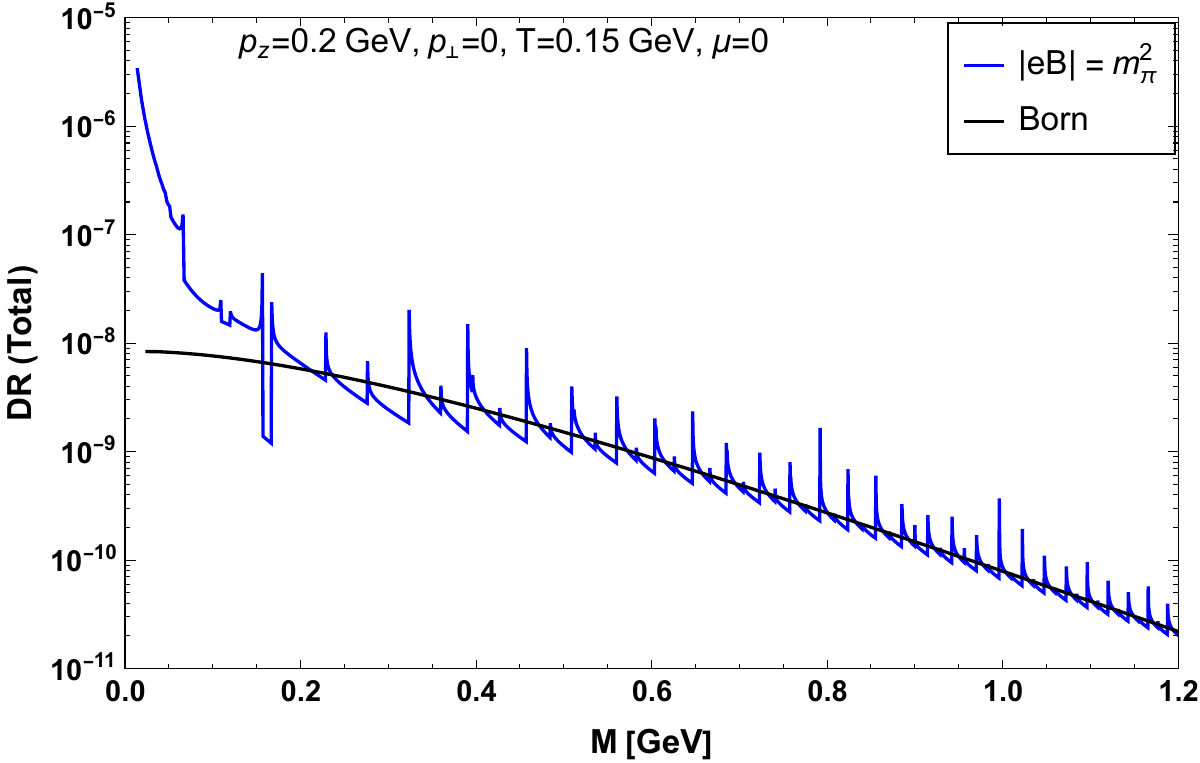}
\includegraphics[scale=0.4]{fig1b.pdf}
\includegraphics[scale=0.4]{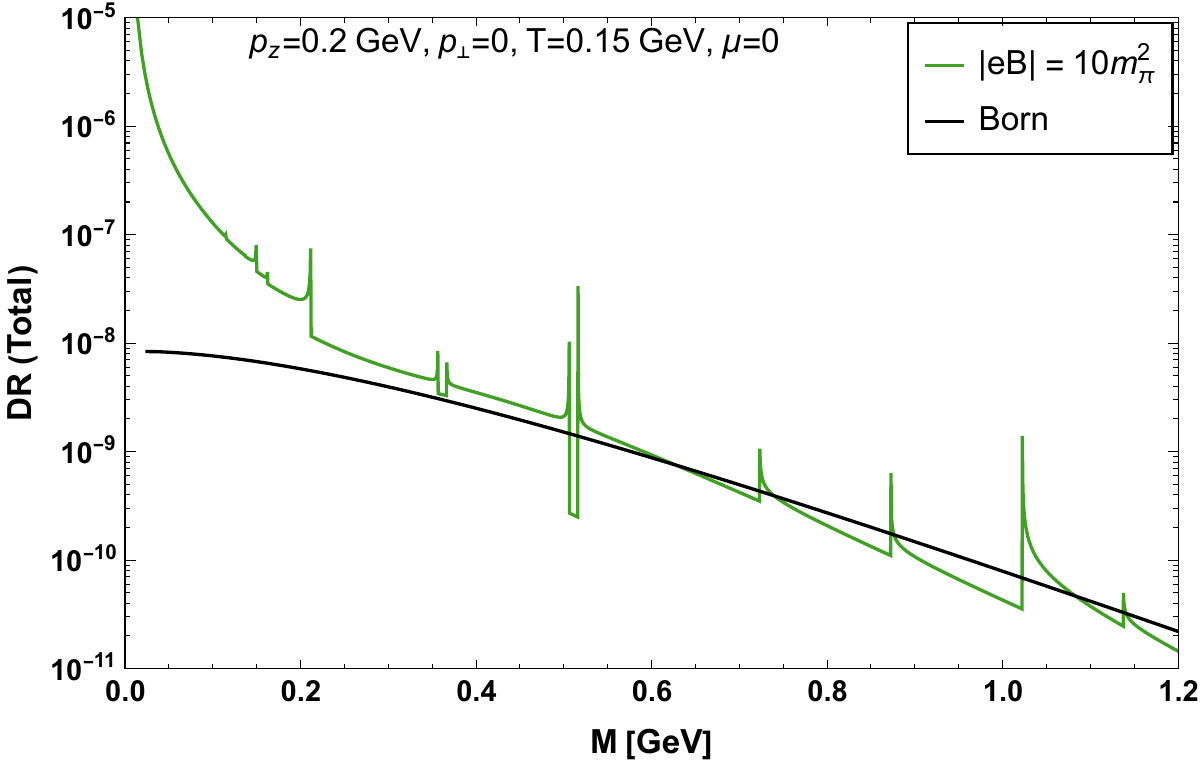}
\includegraphics[scale=0.4]{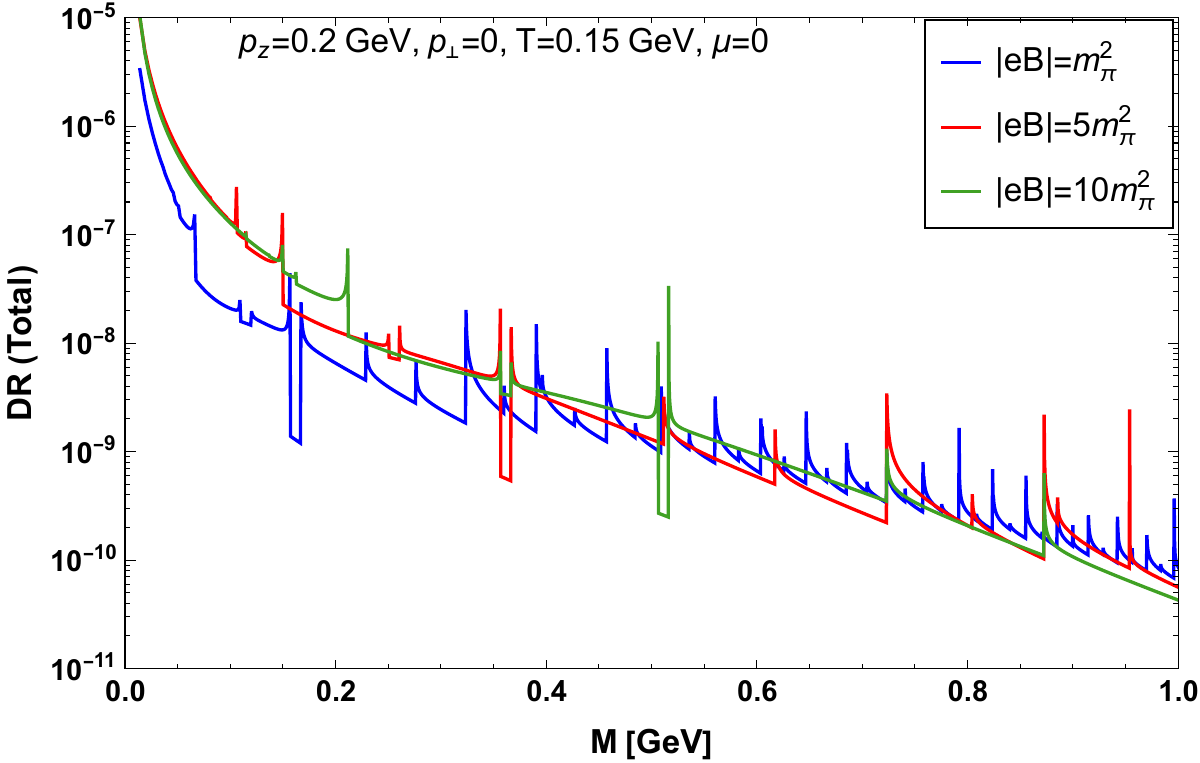}
\caption{(Color online) Plots of dilepton rate as a function of invariant mass for different values of magnetic field. Born rate is also shown for the comparison. The figures are obtained for zero values of $p_{\sperp}$. Please follow the legend and the corresponding text for the values of other parameters. In the last panel we have compared the rates calculated for different values of magnetic field.}
\label{fig:dr_diff_eB}
\end{center}
\end{figure}

The important observation we have from the left panel of the figure~\ref{fig:dr_diff_process} is that there is an enhancement of lepton pair production in presence of $eB$ as compared to the Born rate. The enhancement is observed for a strength of the magnetic field relevant to the ongoing HIC experiments. One can also notice that the enhancement is happening mainly in the regime of lower values of the invariant mass, whereas at comparatively higher invariant masses the effect of the external magnetic field diminishes (except the spikes due to the crossing of the Landau level thresholds) as the DR for the non vanishing magnetic field matches with the Born rate. But most importantly, the enhancement we found is in the detectable range of invariant mass i.e., within the scope of detectors involved in HICs. For arbitrary strengths of the external magnetic field, when one deals with numerous Landau levels (LL) and multiple dilepton production processes, we observe that the contribution from different processes are evident in the nature of the plot of DR as explained in the right panel. 

In the right panel of figure~\ref{fig:dr_diff_process} we decompose the rate coming from different processes and show their contribution to the total rate (vide Eq.~\eqref{eq:DR_total}). The contributions from both quark and antiquark decay processes are the same (which is expected because of lack of any chemical imbalance with $\mu=0$) and mainly dominate the low $M$ region. With increasing invariant mass they start decreasing very fast and the quark-antiquark annihilation process starts overtaking, which becomes the sole contributor in the high enough invariant mass region. At this point some comments are necessary to understand the overlap between the decay and the annihilation processes. We have argued in the previous section~\eqref{sec:cal} that with given other parameters' ($p^2_{\shp}$, $eB$ and $m_f$) values, only one among the three processes shown in Fig.~\ref{fig:dr_diff_process} is allowed for certain values of LLs $\{\ell,n\}$. But when we consider DR at a given invariant mass, we can have overlapping contributions from the decay and the annihilation processes. Because, although the set of LLs that contributes to the annihilation process at a given invariant mass has no elements in common with the set that contributes to the decay processes, there is no condition that doesn't allow dileptons at a given $M$ from both decay and annihilation processes.

To keep track of the total rate for arbitrary $eB$ values, we also show the rate coming solely from lowest Landau level (LLL) in the strong magnetic field approximation calculated in Ref.~\cite{Bandyopadhyay:2016fyd}\footnote{It is to be mentioned here that the expression for the rate in Ref.~\cite{Bandyopadhyay:2016fyd} is incorrect because of the introduction of the effective coupling $(e_e)$ outside the flavour sum. This changes the flavour weights of the rate. We correct that part while comparing with the LLL result. Similar issue is there in Ref.~\cite{Islam:2018sog} which is another LLL exploration.}. It is important to note that in LLL approximation, the contribution comes only from the annihilation process and that even decreases very fast because of having a single LL (see Ref.~\cite{Bandyopadhyay:2016fyd}). This is the reason why both the rates for the LLL approximation and for annihilation process in arbitrary magnetic field start with the same values and then LLL one falls of very sharply, whereas the other one maintains a much higher value because of the involvement of other LLs along with LLL. Similar trend between the LLL and the arbitrary field was also observed for mesonic spectral function~\cite{Chakraborty:2017vvg}.\\

We take the left panel of the Fig.~\ref{fig:dr_diff_process} as a prototype and then use three different values of $eB$ to obtain Fig.~\ref{fig:dr_diff_eB}, while the other parameters are kept at the same values. For the first three panels (top-left, top-right and bottom-left) we have gradually increased the strength of the magnetic field ($m_\pi^2,\;5\,m_\pi^2\;{\rm and}\;10\,m_\pi^2$) encompassing the scenarios relevant to both RHIC and LHC and likewise compare the rate with the Born rate calculated with similar choices of applicable parameters. We observed from Fig.~\ref{fig:dr_diff_process} that the enhancement found in Fig.~\ref{fig:dr_diff_process} for $eB=5\,m_\pi^2$ survive for all three choices of $eB$ that we considered. With higher strength of the magnetic field the enhancement is observed for a higher range of invariant masses and eventually they all become equal to the Born rate as the invariant mass is increased further. In support of these observations we have compared the rates obtained for the above-mentioned values of $eB$ in the last panel (bottom-right), which gives a more conspicuous view. It is also to be mentioned here that values the decay processes dominate in the low $M$ region, whereas the high $M$ region is dominated solely by the annihilation process for all possible choices of $eB$.\\ 

\begin{figure}
\begin{center}
\includegraphics[scale=0.4]{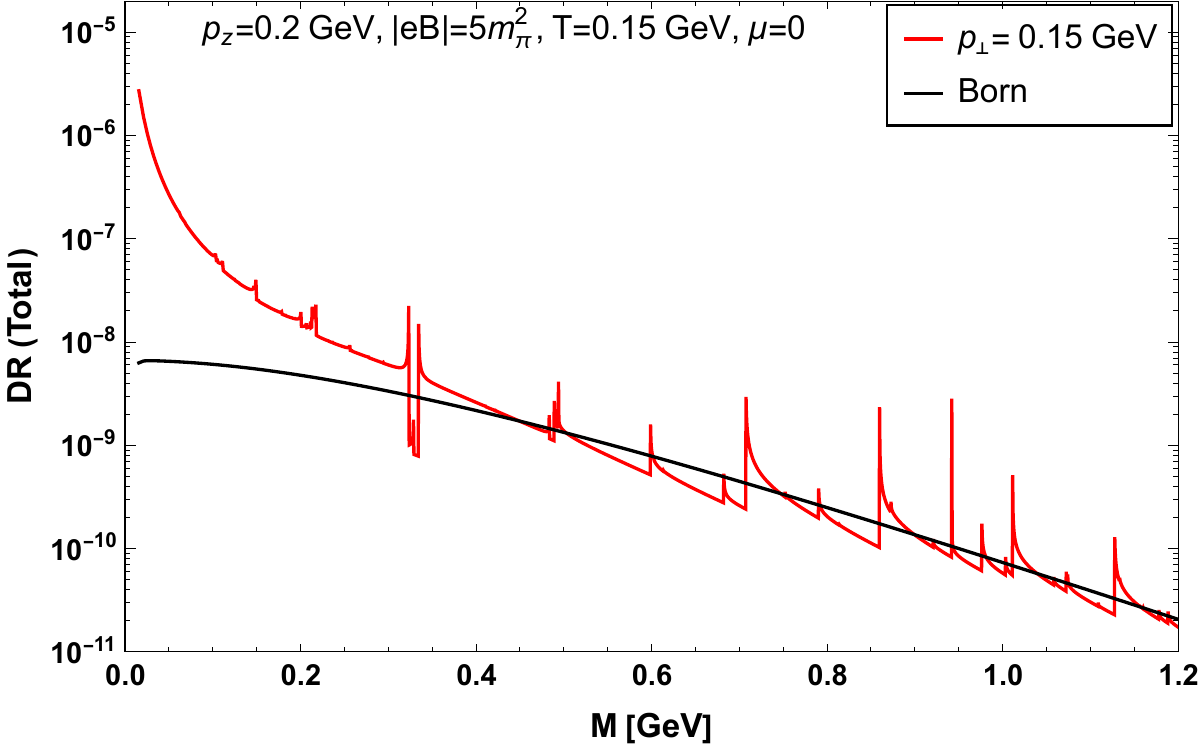} 
\includegraphics[scale=0.4]{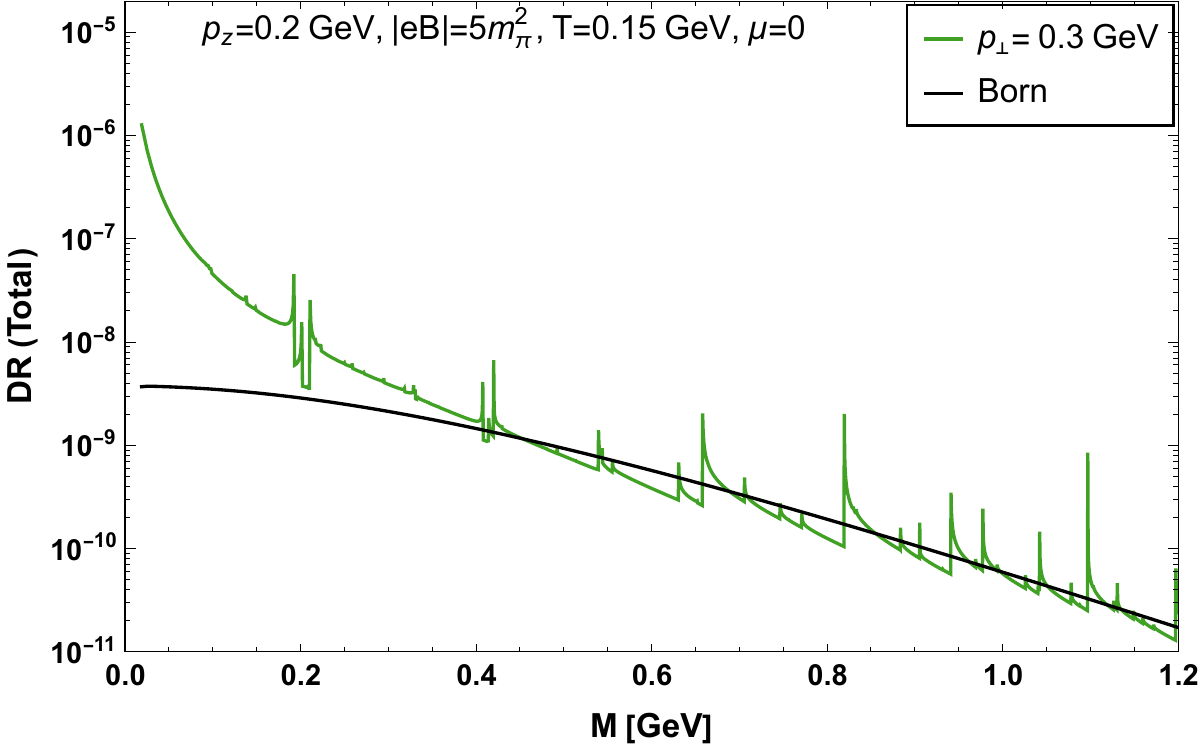}
\includegraphics[scale=0.4]{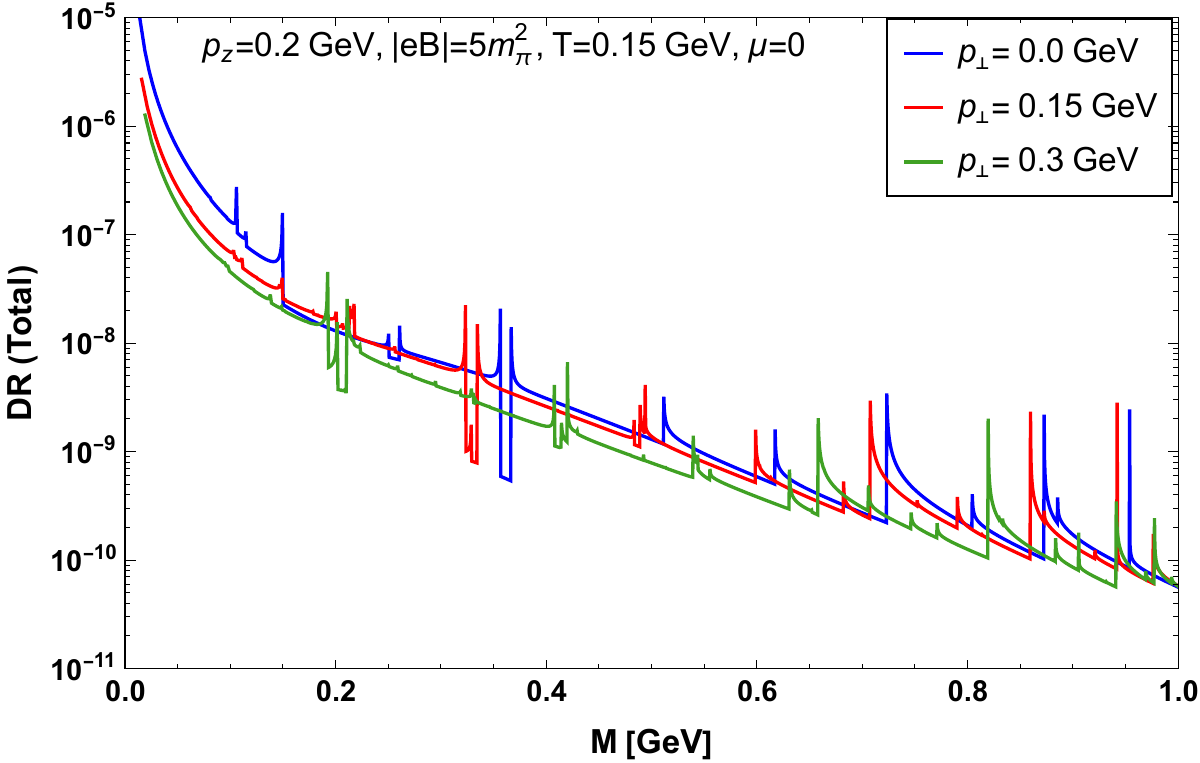}
\caption{(Color online) The top-left and top-right panels show the dilepton rate estimated as a function of invariant mass for two given values of $p_{\sperp}$, i.e. $\{0.15,\,0.3\}$ GeV and given values of $eB$ and $T$. They are also compared with the corresponding born rates. The bottom panel takes a closer view on the effect of $p_{\sperp}$ in the total DR by comparing among three different $p_{\sperp}$ values.}
\label{fig:dr_diff_pT}
\end{center}
\end{figure}

In Fig.~\ref{fig:dr_diff_pT} we show another important feature of the DR made possible by the generalisation of our calculation to arbitrary values of $p_{\sperp}$. It displays the variation of the rate as a function of invariant mass for different values of the momenta along the perpendicular direction of magnetic field. To the best of our knowledge this kind of plot was not previously reported because of the lack of calculation done with simultaneous non-zero values of both $p_z$ and $p_{\sperp}$\footnote{It is to be noted that the rate has been calculated with zero $p_z$ ($p_3$ in their notation) in Ref.~\cite{Sadooghi:2016jyf} and with $p_{\sperp}=0$ in Ref.~\cite{Ghosh:2018xhh}.}. As specimens for nonzero $p_{\sperp}$, we take two different values $\{0.15,\,0.3\}$ GeV for the first two panel (top-left and top-right) in the figure, while the other parameters take the values such as $p_z=0.2$ GeV, $T=0.15$ GeV, $eB=5\, m_\pi^2$ and $\mu=0$ and then compare them with the corresponding Born rates. For both $p_{\sperp}$ values the rate is higher as compared to the Born rate up to certain values of invariant masses. When compared among different $p_{\sperp}$ values themselves, we notice that the rate decreases with increasing $p_{\sperp}$ and $p_{\sperp}=0$ remains the highest among all. It is evident that this happens for both the decay and annihilation processes. All these important features are clearly visible from the lower panel of the figure where we have restricted ourselves for the invariant mass range $0<M<1$ GeV to have a closer view.  

We should mention here a little bit more about the processes for non-zero $p_{\sperp}$. When the transverse momentum $(p_{\sperp}=0)$ of the virtual photon is zero, then a bare quark sitting on the LL, $\ell$ can annihilate only with a bare quark sitting on $\ell-1$, $\ell$, $\ell+1$. Likewise it can decay into a bare quark only with LL $\ell-1$. But in the case of $p_{\sperp}\neq 0$, there is no such restrictions \textemdash\, a bare quark can annihilate with a bare antiquark sitting on any other LL or it can decay into a bare quark sitting on any other LL $n(<\ell)$.

\begin{figure}
\begin{center}
\includegraphics[scale=0.4]{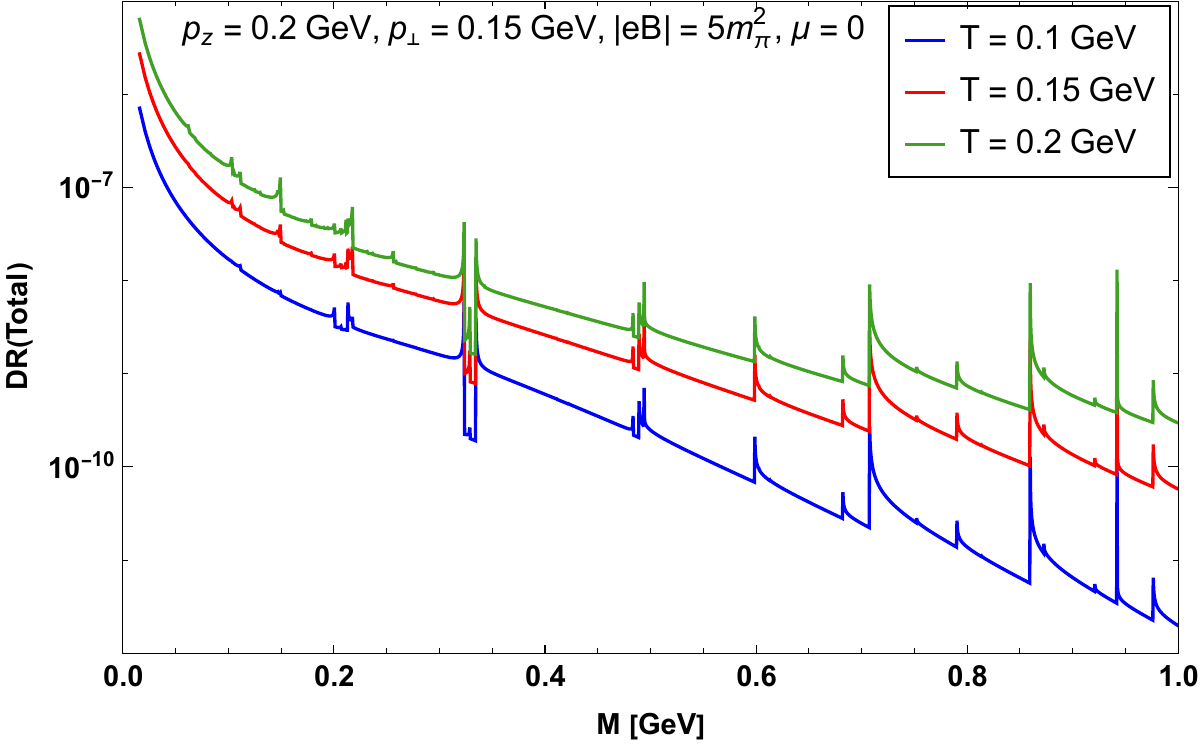}
\caption{(Color online) Dilepton rate is plotted as a function of invariant mass for different values of $T$ at a value of $eB=5 m_\pi^2$. The value of $p_{\sperp}$ is kept at $150$ MeV.}
\label{fig:dr_diff_T}
\end{center}
\end{figure}

In Fig~\ref{fig:dr_diff_T}, We have plotted the DR for different values of temperatures to showcase the impact temperature has on the DR, when other parameters are kept at fixed values. This figure is obtained by keeping $p_\perp = 0.15$ GeV, $\mu=0$, $p_z=0.2$ GeV and $eB=5\,m_\pi^2$. We used three different values of the temperature ($0.1,\,0.2,\,0.3$ GeV), which roughly belong to three different regimes in the QCD phase diagram \textemdash\, below, around and above the phase transition temperatures, respectively. 
It is further observed that the DR is hugely impacted with the change of temperature and it increases when temperature is increased, for both decay and annihilation processes for a given value of $eB$. This behaviour is well known in regard to the Born rate and our finding agrees with that.\\

\begin{figure}
\begin{center}
\includegraphics[scale=0.4]{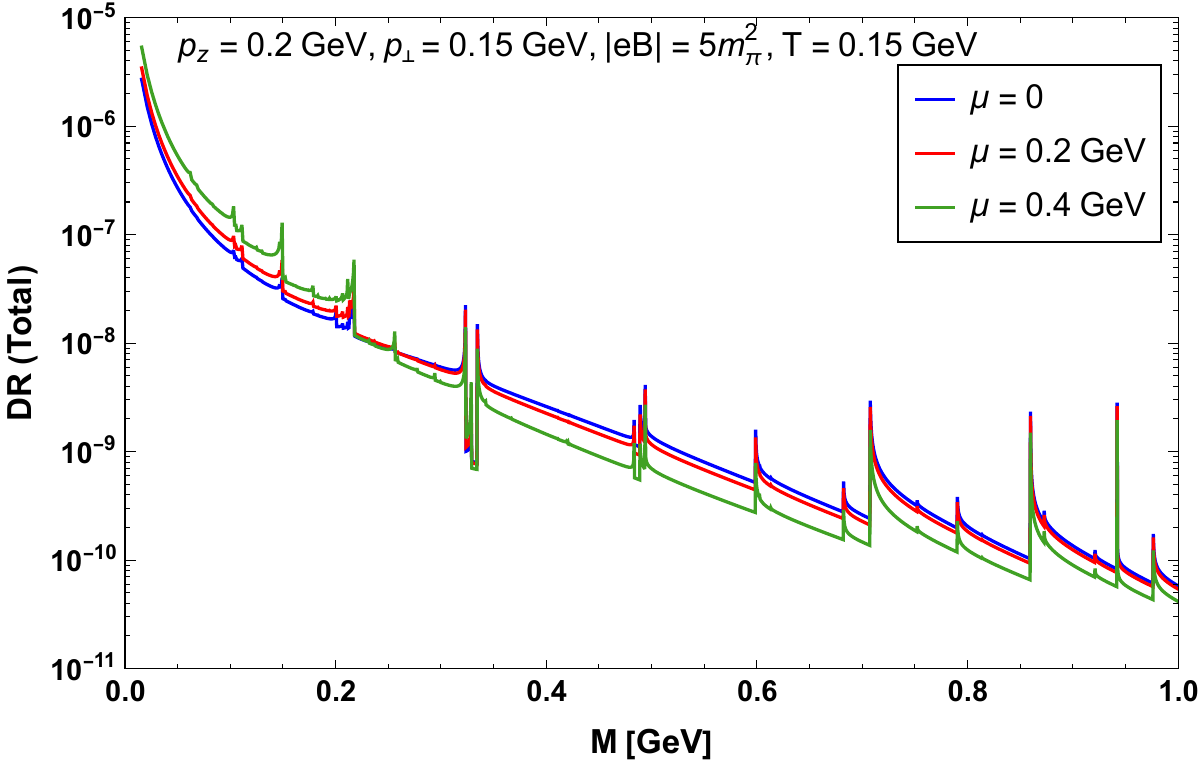}
\includegraphics[scale=0.4]{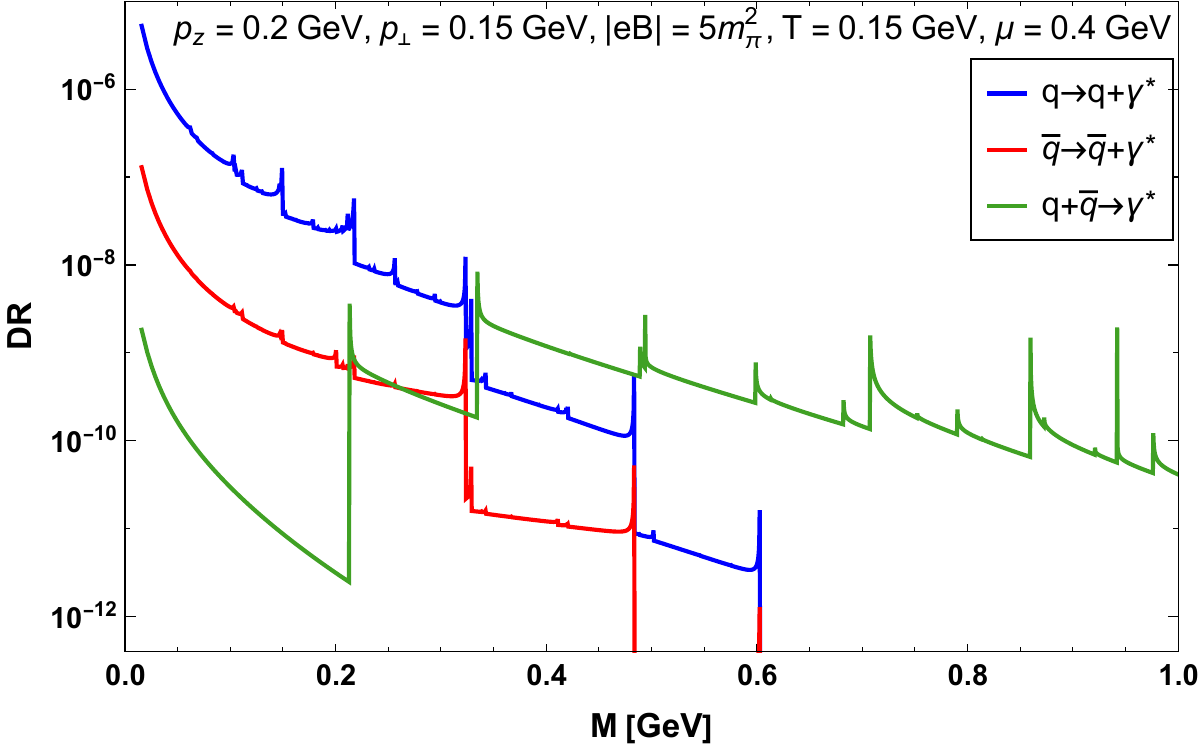}
\caption{(Color online) The effect of the chemical potential on the dilepton rate is exhibited here in the left panel. The plot is drawn for three different values of $\mu$, i.e. $\{0, 0.2,\,0.4\}$ GeV while the $p_{\sperp}$ is kept at $150$ MeV. In the right panel we decompose the same plot where we show quark/antiquark decay and quark-antiquark annihilation processes separately.}
\label{fig:dr_diff_mu}
\end{center}
\end{figure}

After discussing the effects of the temperature, in Fig.~\ref{fig:dr_diff_mu} we have focused on the effects of the chemical potential on the DR, where we show plots for different values of chemical potential. For this plot $p_{\sperp}$ is kept at a value of $0.15$ GeV and the values of other parameters are same as in the previous Fig~\ref{fig:dr_diff_pT}. In the left panel of this Fig~\ref{fig:dr_diff_mu}, we observe that the behavior of the DR at higher invariant mass is similar to that of panel three in Fig~\ref{fig:dr_diff_pT} with the increase of chemical potential. The decrease in the DR with increasing chemical potential is expected and understandable. With higher values of $\mu$, the medium becomes more dense and the probability of a pair of lepton getting lost within it also becomes higher. We also note that for the lower values of the invariant mass, where the decay processes dominate, the behavior is opposite, i.e., with increasing $\mu$, DR increases. From the right panel, where we decompose the rate into different processes, we have another interesting observation. We notice that the rate for the quark decay is higher than that for the antiquark decay because of the nonzero value of $\mu$. This is in sharp contrast to the plot in the right panel of Fig.~\ref{fig:dr_diff_process}, where we have $\mu=0$. Apart from this difference, every other observations remain the same between these two plots. Here, we deliberately refrain ourselves from showing LLL approximated DR plot for non-zero $\mu$ which would have again matched with the annihilation process at the LLL. It is also to be noted that the whole discussion remains valid for vanishing $p_{\sperp}$, as it is evident from Fig.~\ref{fig:dr_diff_pT} that the introduction of $p_{\sperp}$ only brings suppression in the rate.

\subsection{In an effective model scenario}
\label{ssec:res_eff_model}
Next we discuss our results within the NJL model where we have incorporated the IMC effect through a medium dependent scalar coupling. Through this scalar coupling both the MC and the IMC effects in turn get reflected in the effective quark mass $\mathcal{M}_f$, which we have further used in our expression of the DR instead of the bare mass $m_f$.
\begin{figure}
\begin{center}
\includegraphics[scale=0.4]{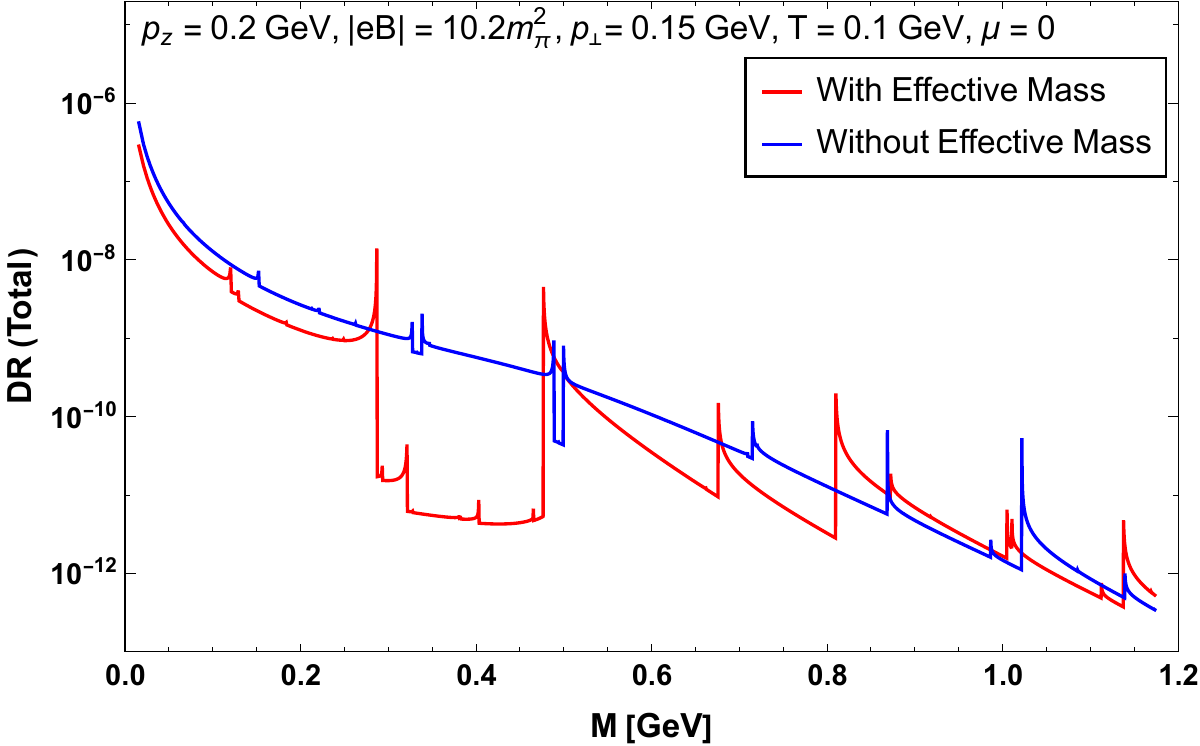}
\caption{(Color online) The comparison between the rates calculated using basic thermal field theory and in the ambit of effective model for the value of $eB=0.2$ GeV. For other details please see the main text.}
\label{fig:dr_eff_mod_com}
\end{center}
\end{figure}

In Fig.~\ref{fig:dr_eff_mod_com}, the total rate has been compared between the two scenarios \textemdash in ordinary finite temperature field theory (FTFT) and in the ambit of effective model, i.e., without and with the effective quark mass $\mathcal{M}_f$, respectively. The magnetic field and temperature we used for this figure are $10.2\,m_\pi^2$ and $0.1$ GeV, respectively, whereas the chemical potential is kept at zero with the different components of the momentum being $p_z=0.2$ and $p_{\sperp}=0.15$ GeV.

We observe from the figure that the rate estimated in the scenario of effective model is suppressed as compared to that calculated using usual FTFT. Not only in terms of the rate, there is another major difference between the two which is the presence of a gap between the rates arising from the decays and annihilation in the effective model scenario. The essential reason of the difference is the mass difference of the quarks between the two scenarios. The quarks simply possess the current masses in the usual FTFT calculation, whereas in the effective model they acquire the effective mass values which depend on the external parameters including the temperature and magnetic field. The mass difference is $\sim 10^2$ at the temperature that we have explored, which causes the gap in the rate calculated using effective model. 
\begin{figure}
\begin{center}
\includegraphics[scale=0.4]{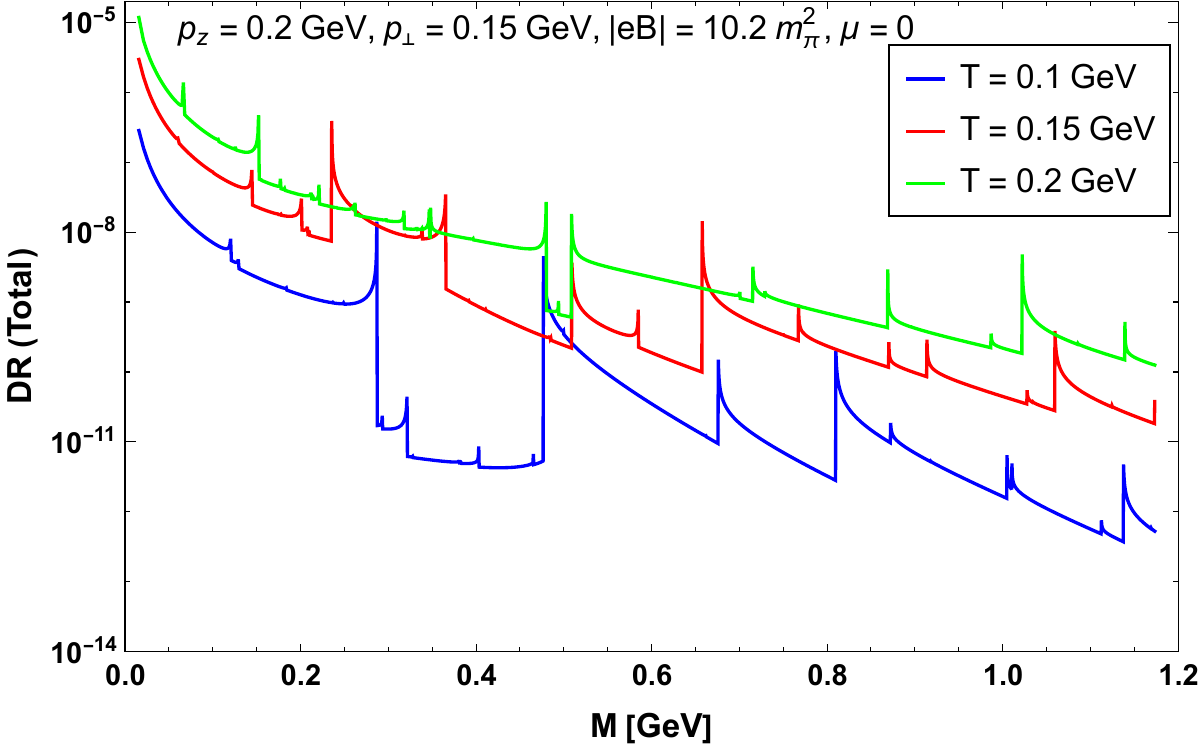}
\includegraphics[scale=0.4]{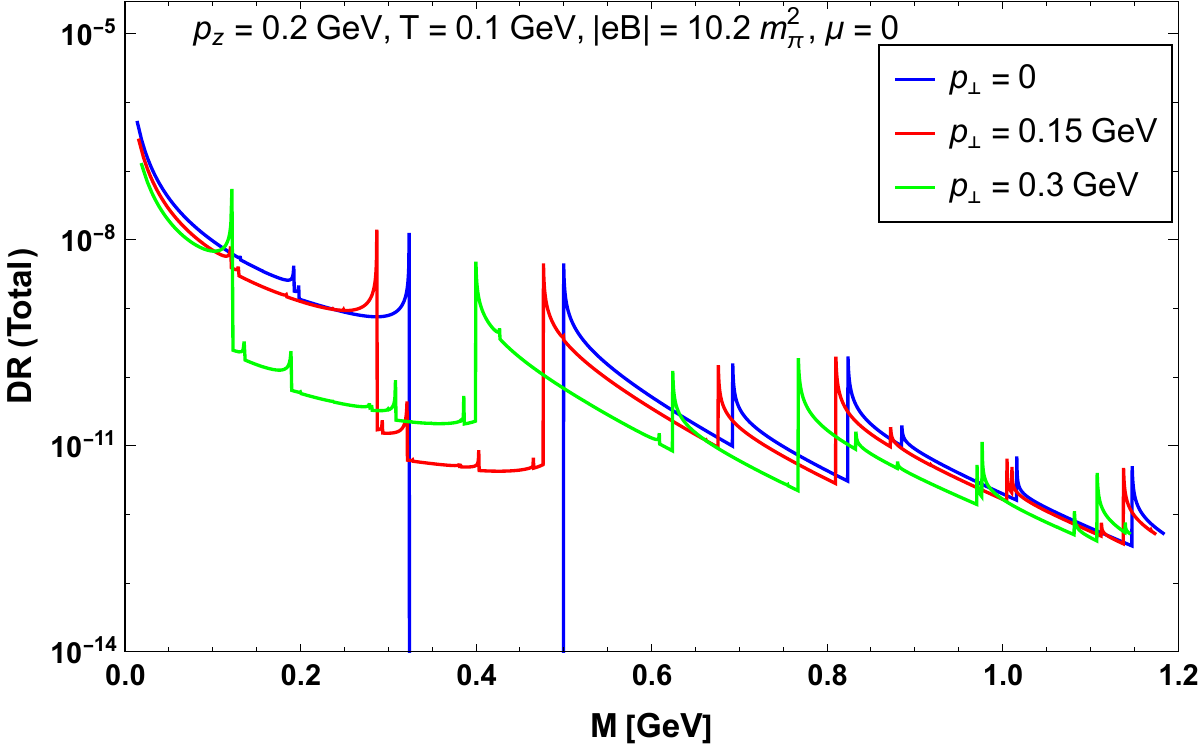}
\caption{(Color online) Left panel: Dilepton rate plotted as a function of invariant mass for three different values of temperatures, in an effective model environment. This plot is analogous to the Fig.~\ref{fig:dr_diff_T} with some fundamental differences. Right panel: Rate plotted for three different values of $p_{\sperp}$ for $T=0.1$ GeV with other parameters being similar to those in the left panel.}
\label{fig:dr_diff_T_eff_mod}
\end{center}
\end{figure}

This effective mass is maximum at zero $T$, which is around $250$ MeV at zero $eB$ in the model that we have applied. With increasing temperature the effective mass starts decreasing and at high enough values, well above the chiral crossover temperature $(T_{CO})$, it attains the current quark mass.  The scenario gets a bit complicated as the external magnetic field is introduced. It causes the effective mass to increase at $T$ lower than the $T_{CO}$, known as magnetic catalysis (MC), and to decrease it around the $T_{CO}$, effectively termed as IMC. Our model incorporates the IMC effect as predicted by LQCD calculation~\cite{Bali:2012zg,Bali:2011qj}. This is a major difference among the present analysis and the previous effective model DR calculations~\cite{Ghosh:2020xwp,Chaudhuri:2021skc} that lack the feature of IMC. The incorporation of IMC will make a crucial difference in terms of our understanding of the rate. We come back to that after stating a few other observations.
\begin{figure}
\begin{center}
\includegraphics[scale=0.4]{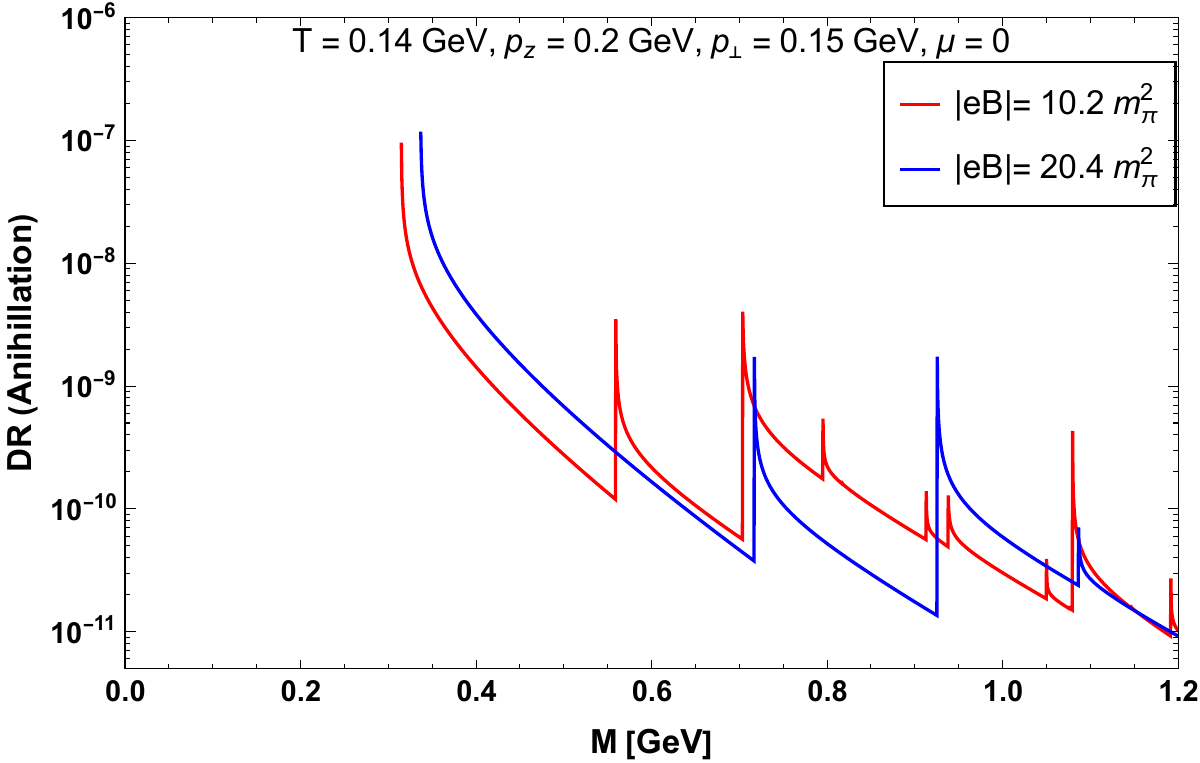}
\includegraphics[scale=0.4]{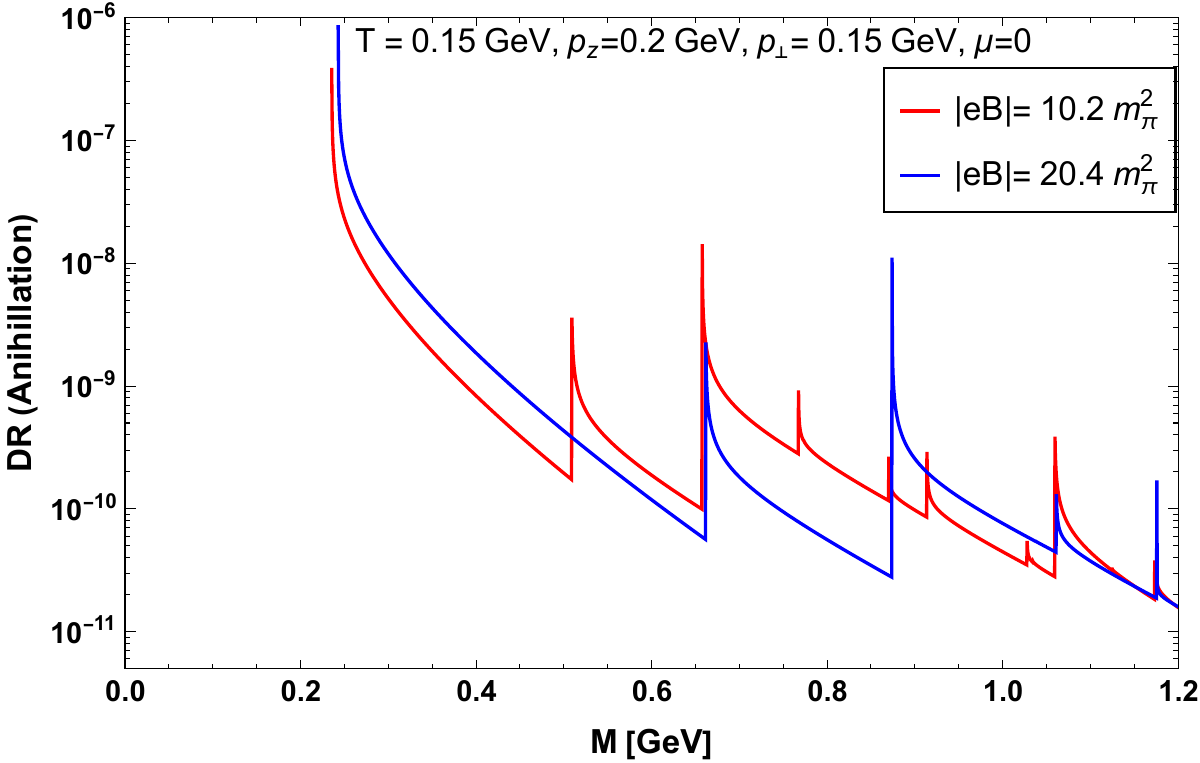}
\includegraphics[scale=0.4]{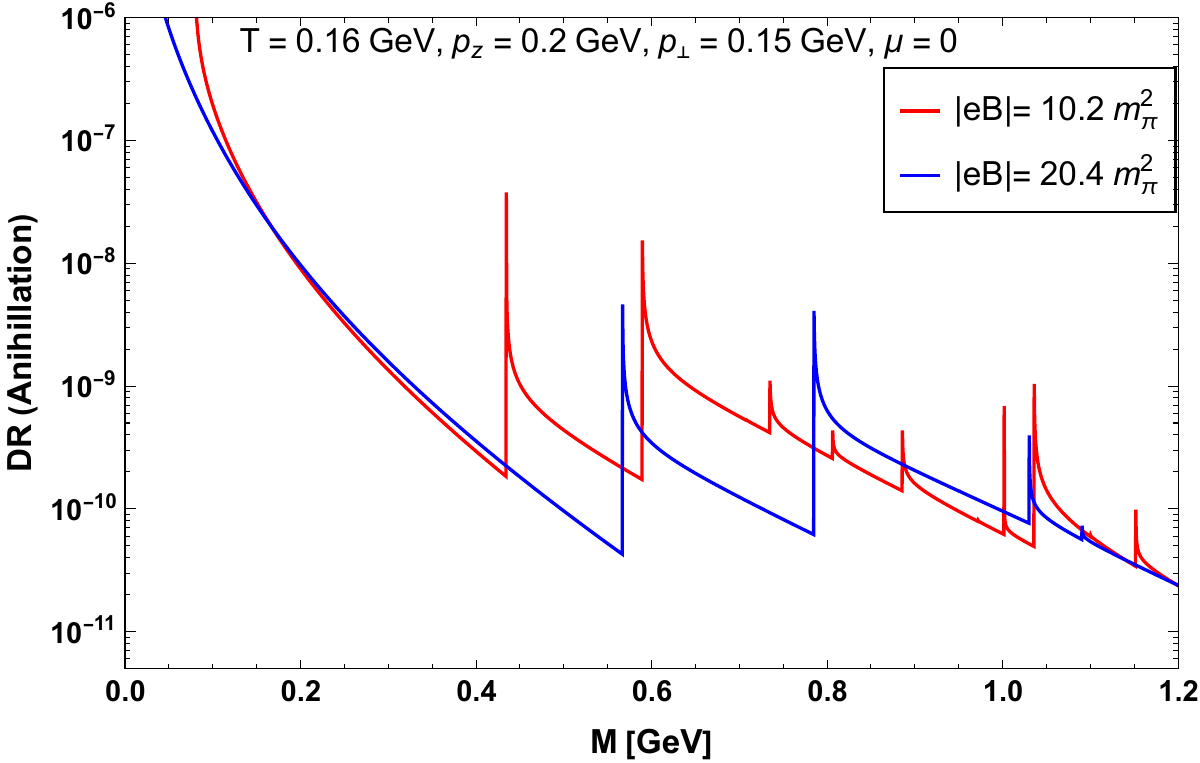}
\caption{(Color online) Comparison of the rates calculated in the effective model for two different values of $eB$ and at three different regions of the QCD phase diagram.}
\label{fig:dr_diff_T_n_eB_eff_mod}
\end{center}
\end{figure}

From the above discussion, it is easy to visualise that if we are at a temperature well below $T_{CO}$ $(\approx 100\, \mathrm{MeV})$ increasing $eB$ will simply increase the rate along with a wider gap between decay and annihilation rates. On the other hand, for a given value of $eB$, let's say $10.2 m_\pi^2$, if we change the temperature the resulting effects are shown in the left panel of Fig.~\ref{fig:dr_diff_T_eff_mod}. There we have plotted the DR for three different values of temperatures in the ambit of effective model. This plot is analogous to the plot given in Fig.~\ref{fig:dr_diff_T} but with some considerable different implications. The three temperatures are chosen to represent three different regions of the QCD phase diagram \textemdash\, below the transition $(0.1\;{\mathrm{GeV}})$, around the transition $(0.15\;{\mathrm{GeV}})$ and above the transition $(0.2\;{\mathrm{GeV}})$.

In Fig.~\ref{fig:dr_diff_T}, the current quarks masses are taken into account and differences among the three coloured lines are only in terms of the temperature. On the other hand, in the figure in discussion (left panel of Fig.~\ref{fig:dr_diff_T_eff_mod}), three different temperatures also give three distinct values of the effective mass which plays important role in the rate as already explained.

For a given $T$ with similar $eB$, the rate gets suppressed in presence of an effective mass, which becomes evident while compared to the Fig.~\ref{fig:dr_diff_T}, particularly the DR lines for $T=0.1$ and $0.2$ GeV. Apart from suppressing the rate in the effective model for a given $T$, the overall effect of temperature remains the same which is to increase the rate with its increasing values. The other important feature being the appearance of the gap in the rate between the decay processes and the annihilation. From this figure it is now evident that this gap depends on the effective mass and increases with its increment. With the increase of $T$ the effective mass decreases and so does the width of the gap and at some certain value of $T$ it disappears. The nature of the gap, particularly the strength of the line connecting the decay and the annihilation processes, also depends on the values of $p_{\sperp}$. This is elaborated in the right panel.  For zero $p_{\sperp}$, the gap is connected by a line with a strength too weak to have any practical impact and also beyond the $y$-axis range of the figure displayed. As $p_{\sperp}$ is increased the connecting line between the gap starts increasing its strength (visible in the figure) and the gap itself widens. It is also noteworthy that the gap shifts towards left with the increase in $p_{\sperp}$ which is understandable as it is plotted as a function of invariant mass.

Finally, we discuss another important feature in the rate, particularly for the annihilation part, through the Fig.~\ref{fig:dr_diff_T_n_eB_eff_mod}. It shows an interesting feature of the annihilation rate calculated around the $T_{CO}$ with two values of $eB$. To be specific we have used three different temperatures for the analysis \textemdash\, $140$, $150$ and $160$ MeV which are just below, around and just above the $T_{CO}$, respectively. Unlike previous effective model exploration, we have the IMC effect incorporated in our model because of which there is a flip in the order of effective masses as we increase $eB$ for those choices of temperature. Thus the effective mass for $eB=20.4\, m_\pi^2$ is higher than that for $eB=10.2\, m_\pi^2$ at $T=140$ MeV. The order flips as we go to $160$ MeV of temperature. This feature reflects at the starting points of the annihilation rates, which reverses their order between the temperature $140$ and $160$ MeV. This is understandable as the higher value of effective mass pushes the rate further towards higher value of invariant masses. We would like to emphasise here that this observation may not be super useful as far as the total rate is concerned, but it is, in our opinion, very crucial in term of our understanding of the behaviour of the rate.

\section{Conclusion}
\label{sec:con}

In this article we have calculated the dilepton rate (DR) from a hot and dense QCD medium in presence of an arbitrary magnetic field $(eB)$. The calculation is performed for simultaneous nonzero values of both the parallel $(p_z)$ and perpendicular $(p_{\sperp})$ components of momentum to the direction of magnetic field. This generalisation, to the best of our knowledge, was not done before and all the previous calculations of DR were carried out under the approximations of either zero parallel~\cite{Sadooghi:2016jyf} or zero perpendicular~\cite{Ghosh:2018xhh} components of the momentum. 

We have also decomposed the rate into different processes that it consists of. It is to be noted that in absence of magnetic field it is only the annihilation of quark-antiquark which is kinematically allowed; other processes are prohibited. But in presence of $eB$ (except when one uses the LLL approximation) the decay processes of both quark and antiquark also become kinematically allowed along with the annihilation. This is the most significant change as we switch from zero to non-zero $eB$ scenario. Physically the decay processes can be understood in the following way: In the presence of a magnetic field, a bare quark or anti-quark can lose energy by going from a higher Landau level (LL) to a lower one by emitting a virtual photon which can further decay into a $l\bar{l}$ pair.

As our first major finding, we observe that the DR in presence of non-zero $eB$ is enhanced as compared to the Born rate in the one loop order. The enhancement is significant, particularly at the lower range of the invariant mass region, where the contribution is maximum from the decay processes. As we go along the direction of increasing invariant mass, the rate from the decay processes start diminishing and the total rate is mainly controlled by the annihilation process alone, roughly above the invariant mass of $400$ MeV. Around the same invariant mass, the rate also starts to merge with the Born rate. The enhancement in the low invariant mass region is definitely encouraging but needs to be considered carefully as this region is related to the late times in the evolution of the fireball for which the strength of the magnetic field is expected to be much weaker.

As permitted by our generalised framework, we also study the effect of nonzero perpendicular component of the momentum on the DR. We observe that for a given value of magnetic field, the nonzero $p_{\sperp}$ suppresses the DR as compared to zero $p_{\sperp}$ and the suppression is all along the range of the invariant mass. Although, in that case as well, we found sufficient enhancement in the rate as compared to Born rate. Subsequently we have shown all the following plots with finite values of $p_{\sperp}$ to showcase the novelty of the present calculation.

We also explored the effects of other parameters like temperature and chemical potential on the rate. As is well known, the temperature enhances the rate for a given value of $eB$, whereas the chemical potential decreases it. With the introduced chemical imbalance we decomposed the rate into different processes. We found that the contribution from the quark decay is higher than that from the anti-quark decay, which is expected.

As another novel feat of the present study, we have explored the DR within an effective NJL model scenario. Here we have incorporated both the MC and IMC effects through the effective quark mass by using a medium dependent coupling constant and observed its effect on the DR, which again to the best of our knowledge has not been done before. We have discussed about the effect of the medium dependent effective quark mass on the DR and subsequent occurrence of a mass gap between the decay and the annihilation processes for certain values of the temperature and the external magnetic field.  

The interesting behaviour of the gap between the decay and annihilation rates at temperatures lower than the phase transition temperature has been explored in detail. Particularly, its behaviour as a function of $p_{\sperp}$ is investigated minutely. The gap widens and shifts towards left, whereas the line connecting the gap increases its strength as $p_{\sperp}$ is increased. We have also explicitly noticed the interesting effects of IMC on the rate, specifically on the annihilation rate, near the crossover temperature.

As an outlook, it would be really interesting to explore the space-time evolved dilepton spectra in presence of an arbitrary external magnetic field with the help of the currently obtained dilepton production rate. The DR evaluated in the present study will be considered as an input in the natural framework of the relativistic hydrodynamics to perform the space time evolution and hence obtain the dilepton spectra, i.e., 
\begin{align}
    \frac{dN}{dM} = \int d^4X \frac{d^3p M}{p_0} \left(\frac{dN}{d^4X d^4P}\right).
\end{align} 
Although one has to be really careful while choosing the magnetic field profile required for the space time integration which will greatly impact the numerical procedure in our opinion. Evaluation of the dilepton spectra would then allow the present study to be tested by calculating some directly experimentally viable observables.

\section{Acknowledgements}
\label{sec:acknow}
The authors are thankful to Najmul Haque for his valuable comments and suggestions. We are really thankful to Xinyang Wang and Igor A. Shovkovy for pointing out an important issue with the rate expression in the older version. Help from Kun Xu on running one programme on the cluster of the University of Chinese Academy of Sciences, Beijing is much appreciated.  A.D. acknowledges Department of Atomic Energy, India. He is particularly grateful to his parents for providing him with a cutting edge laptop which was very useful for performing the heavy numerical analysis. A.B. acknowledges the support from Guangdong Major Project of Basic and Applied Basic Research No. 2020B0301030008,  China and the postdoctoral research fellowship from the Alexander von Humboldt Foundation, Germany. C.A.I. would like to acknowledge the financial support by the Chinese Academy of Sciences President's International Fellowship Initiative under Grant No. 2020PM0064 and partial help by the Fundamental Research Funds for the Central Universities, China.

\appendix
\section{Analytic structure} 
\label{app:ana_struc}
In this section we discuss the details of the analytic structure of the imaginary part of the self-energy. The imaginary part of the self energy $\textsf{Im}\Pi^{\mu}_{f,\mu}(p_0,\bm{p}_{\shp},p_z)$ has itself a very rich analytic structure as a function of $p_0$ (keeping all other parameters held at a fixed value) in the presence of a hot magnetised medium.  The branch cuts are situated in the regions where delta functions can give nonzero contributions. There are two kinds of branch cuts that appear in the imaginary part of the self energy. One is called unitary cut and the other is called Landau cut. Unitary cuts are present in the analytic structure of the self energy for zero as well as nonzero temperature whereas Landau cut is present only at nonzero temperature. 
\subsection{Unitary cuts} 
\label{app:unitary_cuts}
In this subsection, we discuss unitary cuts in details. When $p_0=\pm(E_{f,\ell,k}+E_{f,n,q})$ is satisfied, the delta functions with $s_1=s_2=\pm 1$ give contributions.
Now let us look at the following term
\begin{align}
(E_{f,\ell,k}+E_{f,n,q})^2 &= E_{f,\ell,k}^2+E_{f,n.q}^2+2 E_{f,\ell,k}E_{f,n,q} \nn
&= k_z^2+m^2_{f,\ell}+(p_z-k_z)^2+m^2_{f,n}+2 E_{f,\ell,k} E_{f,n,q} \nn
&= p^2_z-2 k_z (p_z-k_z)+(m_{f,\ell}+m_{f,n})^2 - 2m_{f,\ell}m_{f,n}+2 E_{f,\ell,k} E_{f,n,q} \nn
&= p^2_z+(m_{f,\ell}+m_{f,n})^2 + 2 ( E_{f,\ell,k} E_{f,n,q}- k_z (p_z-k_z)-m_{f,\ell}m_{f,n}).
\label{eq:Ek+Eq}
\end{align}
Firstly,  we need to make use of Schwartz identity which states that for any two $N$ dimensional vectors $A$ and $B$, we have 
\begin{align}
\left(\sum_{i=1}^{N} A_i^2\right)\left(\sum_{i=1}^{N} B_i^2\right)\geq \left(\sum_{i=1}^N A_iB_i\right)^2.\label{eq:schwartz_inequality}
\end{align}

In this case, we take 
$A=(k_z,m_{f,\ell})$ and $B=(p_z-k_z,m_{f,n})$. It leads to
\begin{align}
E_{f,\ell,k}^2E_{f,n,q}^2\geq (k_z(p_z-k_z)+m_{f,\ell}m_{f,n})^2. 
\end{align}
Since $E_{f,\ell,k}$,$E_{f,n,q} > 0$, we have,
\begin{align}
E_{f,\ell,k}E_{f,n,q}\geq k_z(p_z-k_z)+m_{f,\ell}m_{f,n}\,.
\label{eq:uni_sch_ineq}
\end{align} 
Thus from Eqs.~\eqref{eq:Ek+Eq} and \eqref{eq:uni_sch_ineq}, we get
\begin{align}
(E_{f,\ell,k}+E_{f,n,q})^2 \geq p^2_z+(m_{f,\ell}+m_{f,n})^2,
\end{align}
which eventually leads to
\begin{align}
E_{f,\ell,k} + E_{f,n,q} \geq \sqrt{p^2_z+(m_{f,\ell}+m_{f,n})^2}\,.
\end{align}
From the above equation it is evident that 
\begin{align} 
p_0 - E_{f,\ell,k} - E_{f,n,q} &\leq p_0 - \sqrt{p^2_z+(m_{f,\ell}+m_{f,n})^2}\;\; {\rm and}\\
p_0 + E_{f,\ell,k} + E_{f,n,q} &\geq p_0 + \sqrt{p^2_z+(m_{f,\ell}+m_{f,n})^2}\,.
\end{align}
Now from the above inequality we see that the argument of $\delta(p_0+E_{f,\ell,k}+E_{f,n,q})$ will always be greater than zero for $k_z\in (-\infty,\infty)$ and for any $\ell,n\geq 0$. Thus we can drop this term.
Finally for the annihilation of process, we impose this condition by multiplying an overall factor of $\Theta\left(p_{\shp}^2-\left(m_{f,\ell}+m_{f,n}\right)^2\right)$.

Note that for large  $k_z$, $(p_0-E_{f,\ell,k}-E_{f,n,q})$ is unbounded from below.  In fact,  it goes as
\begin{align}
(p_0 - E_{f,\ell,k}-E_{f,n,q}) \xrightarrow{\text{large }\,|k_z|} -2|k_z|.
\end{align} 
Finally,  we must mention that in the complex $p_0$ plane, there are two unitary cuts that can be chosen to run along the $\textsf{Re}(p_0)$ in the region 
\begin{align}
\left(\left.-\infty\,\,,\,\,-\sqrt{p^2_z+\left(m_{f,\ell}+m_{f,n}\right)^2}\right.\right]\qquad\text{and}\qquad \left[\left.\sqrt{p^2_z+\left(m_{f,\ell}+m_{f,n}\right)^2}\,\,,\,\,\infty\right.\right).
\end{align} 
\subsection{Landau cut} 
\label{app:landau_cut}
In this subsection, we talk about the Landau cuts in details. When $p_0 = \pm(E_{f,\ell,k}-E_{f,n,q})$ is satisfied, the delta functions with $s_1 =-s_2=1$ and $s_1=-s_2=-1$ give  contributions. So, as before, we first compute
\begin{align}
(E_{f,\ell,k}-E_{f,n,q})^2 &= E_{f,\ell,k}^2+E_{f,q,n}^2-2 E_{f,\ell,k}E{f,q,n} \nn
&= k_z^2+m^2_{f,\ell}+(k_z-p_z)^2+m^2_{f,n}-2 E_{f,\ell,k} E_{f,n,q} \nn
&= p^2_z+2 k_z (k_z-p_z)+(m_{f,\ell}-m_{f,n})^2 + 2m_{f,\ell}m_{f,n}-2 E_{f,\ell,k} E_{f,n,q} \nn
&= p^2_z+(m_{f,\ell}-m_{f,n})^2 - 2 ( E_{f,\ell,k} E_{f,n,q}- k_z (k_z-p_z)-m_{f,\ell}m_{f,n}).
\label{eq:Ek-Eq}
\end{align}
Using Schwartz-inequality given in Eq.~\eqref{eq:schwartz_inequality} with $A=(k_z,m_{f,\ell})$ and $B=(k_z-p_z,m_{f,n})$, we arrive at
\begin{align}
E_{f,\ell,k}E_{f,n,q} \geq k_z (k_z-p_z)+m_{f,\ell}m_{f,n}\,. \label{eq:Landau_sch_ineq}
\end{align}
Likewise   we see from Eqs.~\eqref{eq:Ek-Eq} and \eqref{eq:Landau_sch_ineq} that
\begin{align}
&(E_{f,n,q}-E_{f,\ell,k})^2 \leq p^2_z+(m_{f,\ell}-m_{f,n})^2.
\end{align}
Or in other words
\begin{align}
-\sqrt{p^2_z+(m_{f,\ell}-m_{f,n})^2} \leq E_{f,\ell,k}-E_{f,n,q} \leq \sqrt{p^2_z+(m_{f,\ell}-m_{f,n})^2}.
\end{align}
Thus, it is clear that for real $k_z\in(-\infty,\infty)$, $p_{0}\mp E_{f,\ell,k}\pm E_{f,n,q}$ must lie inside the interval 
\begin{align}
\left[p_0-\sqrt{p^2_z+(m_{f,\ell}-m_{f,n})^2}\,\,,\,\,p_0+\sqrt{p^2_z+(m_{f,\ell}-m_{f,n})^2}\right]. 
\end{align} 
The theta function $\Theta\left(p_{\shp}^2-\left(m_{f,\ell}-m_{f,n}\right)^2\right)$ in Eq.~\eqref{eq:dirac_delta_simplified} enforces this fact for the decay processes.
Now we find the limiting value to which $E_{f,\ell,k}-E_{f,n,q}$ goes when $|k_z|$ is large.  We have
\begin{align}
E_{f,\ell,k} - E_{f,n,q}&= \sqrt{k_z^2+m^2_{f,\ell}}-\sqrt{(k_z-p_z)^2+m^2_{f,n}} \nn
&=\sqrt{k_z^2+m^2_{f,\ell}} - \sqrt{k^2_z-2k_zp_z+p_z^2+m_{f,n}^2}\nn
&=\sqrt{k_z^2}\left[\sqrt{1+\frac{m^2_{f,\ell}}{k^2_z}}-\sqrt{1-2\,\frac{p_z}{k_z}+\frac{p^2_z}{k^2_z}+\frac{m^2_{f,n}}{k^2_z}}\right]\nn
&=\sqrt{k_z^2}\left[\left(1+\frac{m^2_{f,\ell}}{k^2_z}+\cdots\right)-\left(1-\frac{p_z}{k_z}+\frac{p^2_z}{2k^2_z}+\frac{m^2_{f,n}}{2k^2_z}+\cdots\right)\right]\nn
&\simeq\sqrt{k_z^2}\left[1-\left(1-\frac{p_z}{k_z}\right)\right] \nn
&\simeq \textsf{sgn}(k_z)p_z\,.
\end{align}
Thus the asymptotic behaviour of the arguments of delta functions are given as follows
\begin{align}
p_0\pm E_{f,\ell,k} \mp E_{f,n,q} \xrightarrow{\textmd{large  }|k_z|} p_0 \pm \textsf{sgn}(k_z)p_z\,.
\end{align} 
Similarly, we see that Landau cut runs along $\textsf{Re}(p_0)$ axis in the region  
\begin{align}
\left[-\sqrt{p^2_z+\left(m_{f,\ell}-m_{f,n}\right)^2}\,\,,\,\,\sqrt{p^2_z+\left(m_{f,\ell}-m_{f,n}\right)^2}\right].
\end{align} 
\section{Integral over the perpendicular momenta}
\label{app:perp_intg}
In this section, we derive an alternative method to the Ref.~\cite{Wang:2020dsr} to evaluate the perpendicular integrals.
To do so, we consider the integration 
\begin{align}
\mathcal{I}_{f,\alpha}(p^2_{\sperp})\equiv\int\frac{d^2k_{\perp}}{(2\pi)^2}\exp\left(-\frac{k_{\sperp}^2+q_{\sperp}^2}{|q_{\sF}B|}\right)(k.q)_{\sperp}^{\alpha}\exp\left(-\frac{1}{1-t}\frac{2k_{\sperp}^2}{|q_{\sF}B|}\right)\exp\left(-\frac{u}{1-u}\frac{2q_{\sperp}^2}{|q_{\sF}B|}\right)\frac{1}{(1-t)^{\alpha}}\frac{1}{(1-u)^{\alpha}}\,. 
\label{eq:app:perp_gen}
\end{align}
As mentioned before, we perform the integration for only $\alpha =0\;{\rm and}\;1$, which gives
\begin{align}
\mathcal{I}_{f,0}(p^2_{\sperp}) &= \frac{|q_{\sF}B|}{8\pi}\exp\left(-\frac{p_{\sperp}^2}{2|q_{\sF}B|}\frac{(1+t)(1+u)}{1-tu}\right)\frac{1}{1-tu}\;\; {\rm and}
\label{eq:app:If0}\\
\mathcal{I}_{f,1}(p^2_{\sperp}) &= \frac{|q_{\sF}B|^2}{16\pi}\exp\left(-\frac{p_{\sperp}^2}{2|q_{\sF}B|}\frac{(1+t)(1+u)}{1-tu}\right)\frac{1}{(1-tu)^2}\left[1-\frac{(1+t)(1+u)}{1-tu}\frac{p_{\sperp}^2}{2|q_{\sF}B|}\right], \label{eq:app:If1}
\end{align}
respectively. Using Eq.~\eqref{eq:Laguerre_generating},  Eq.~\eqref{eq:app:If0} and Eq.~\eqref{eq:app:If1}, we can rewrite the identities as,
\begin{align}
\sum_{\ell,n=0}^{\infty}\mathcal{J}^{(0)}_{f,\ell,n}(p_{\sperp}^2)t^{\ell}u^n &= \frac{|q_{\sF}B|}{8\pi}\exp\left(-\frac{p_{\sperp}^2}{2|q_{\sF}B|}\frac{(1+t)(1+u)}{1-tu}\right)\frac{1}{1-tu}\;\; {\rm and}\label{eq:app:If0_final}\\
\sum_{\ell,n=0}^{\infty}\mathcal{J}^{(1)}_{f,\ell,n}(p_{\sperp}^2)t^{\ell}u^n &= \frac{|q_{\sF}B|^2}{16\pi}\exp\left(-\frac{p_{\sperp}^2}{2|q_{\sF}B|}\frac{(1+t)(1+u)}{1-tu}\right)\frac{1}{(1-tu)^2}\left[1-\frac{(1+t)(1+u)}{1-tu}\frac{p_{\sperp}^2}{2|q_{\sF}B|}\right], 
\label{eq:app:If1_final}
\end{align}
respectively. Now, using the Taylor expansion of the right hand side of Eq.\eqref{eq:app:If0_final} and Eq.~\eqref{eq:app:If1_final} about $t=u=0$ and comparing $t^{\ell}u^{n}$ terms in the expressions, we get our desired perpendicular momentum integrals as shown in Eq.~\eqref{eq:J0} and Eq.~\eqref{eq:J1}.

\bibliography{reference}

\begin{thebibliography}{46}%
\makeatletter
\providecommand \@ifxundefined [1]{%
 \@ifx{#1\undefined}
}%
\providecommand \@ifnum [1]{%
 \ifnum #1\expandafter \@firstoftwo
 \else \expandafter \@secondoftwo
 \fi
}%
\providecommand \@ifx [1]{%
 \ifx #1\expandafter \@firstoftwo
 \else \expandafter \@secondoftwo
 \fi
}%
\providecommand \natexlab [1]{#1}%
\providecommand \enquote  [1]{``#1''}%
\providecommand \bibnamefont  [1]{#1}%
\providecommand \bibfnamefont [1]{#1}%
\providecommand \citenamefont [1]{#1}%
\providecommand \href@noop [0]{\@secondoftwo}%
\providecommand \href [0]{\begingroup \@sanitize@url \@href}%
\providecommand \@href[1]{\@@startlink{#1}\@@href}%
\providecommand \@@href[1]{\endgroup#1\@@endlink}%
\providecommand \@sanitize@url [0]{\catcode `\\12\catcode `\$12\catcode
  `\&12\catcode `\#12\catcode `\^12\catcode `\_12\catcode `\%12\relax}%
\providecommand \@@startlink[1]{}%
\providecommand \@@endlink[0]{}%
\providecommand \url  [0]{\begingroup\@sanitize@url \@url }%
\providecommand \@url [1]{\endgroup\@href {#1}{\urlprefix }}%
\providecommand \urlprefix  [0]{URL }%
\providecommand \Eprint [0]{\href }%
\providecommand \doibase [0]{http://dx.doi.org/}%
\providecommand \selectlanguage [0]{\@gobble}%
\providecommand \bibinfo  [0]{\@secondoftwo}%
\providecommand \bibfield  [0]{\@secondoftwo}%
\providecommand \translation [1]{[#1]}%
\providecommand \BibitemOpen [0]{}%
\providecommand \bibitemStop [0]{}%
\providecommand \bibitemNoStop [0]{.\EOS\space}%
\providecommand \EOS [0]{\spacefactor3000\relax}%
\providecommand \BibitemShut  [1]{\csname bibitem#1\endcsname}%
\let\auto@bib@innerbib\@empty
\bibitem [{\citenamefont {Muller}(1985)}]{Muller:1983ed}%
  \BibitemOpen
  \bibfield  {author} {\bibinfo {author} {\bibfnamefont {B.}~\bibnamefont
  {Muller}},\ }\href@noop {} {\bibfield  {journal} {\bibinfo  {journal} {Lect.
  Notes Phys.}\ }\textbf {\bibinfo {volume} {225}},\ \bibinfo {pages} {1}
  (\bibinfo {year} {1985})}\BibitemShut {NoStop}%
\bibitem [{\citenamefont {Heinz}\ and\ \citenamefont
  {Jacob}(2000)}]{Heinz:2000bk}%
  \BibitemOpen
  \bibfield  {author} {\bibinfo {author} {\bibfnamefont {U.~W.}\ \bibnamefont
  {Heinz}}\ and\ \bibinfo {author} {\bibfnamefont {M.}~\bibnamefont {Jacob}},\
  }\href@noop {} {\  (\bibinfo {year} {2000})},\ \Eprint
  {http://arxiv.org/abs/nucl-th/0002042} {arXiv:nucl-th/0002042} \BibitemShut
  {NoStop}%
\bibitem [{\citenamefont {Pasechnik}\ and\ \citenamefont
  {\v{S}umbera}(2017)}]{Pasechnik:2016wkt}%
  \BibitemOpen
  \bibfield  {author} {\bibinfo {author} {\bibfnamefont {R.}~\bibnamefont
  {Pasechnik}}\ and\ \bibinfo {author} {\bibfnamefont {M.}~\bibnamefont
  {\v{S}umbera}},\ }\href {\doibase 10.3390/universe3010007} {\bibfield
  {journal} {\bibinfo  {journal} {Universe}\ }\textbf {\bibinfo {volume} {3}},\
  \bibinfo {pages} {7} (\bibinfo {year} {2017})},\ \Eprint
  {http://arxiv.org/abs/1611.01533} {arXiv:1611.01533 [hep-ph]} \BibitemShut
  {NoStop}%
\bibitem [{\citenamefont {Fukushima}(2012)}]{Fukushima:2011jc}%
  \BibitemOpen
  \bibfield  {author} {\bibinfo {author} {\bibfnamefont {K.}~\bibnamefont
  {Fukushima}},\ }\href {\doibase 10.1088/0954-3899/39/1/013101} {\bibfield
  {journal} {\bibinfo  {journal} {J. Phys. G}\ }\textbf {\bibinfo {volume}
  {39}},\ \bibinfo {pages} {013101} (\bibinfo {year} {2012})},\ \Eprint
  {http://arxiv.org/abs/1108.2939} {arXiv:1108.2939 [hep-ph]} \BibitemShut
  {NoStop}%
\bibitem [{\citenamefont {Yagi}\ \emph {et~al.}(2005)\citenamefont {Yagi},
  \citenamefont {Hatsuda},\ and\ \citenamefont {Miake}}]{Yagi:2005yb}%
  \BibitemOpen
  \bibfield  {author} {\bibinfo {author} {\bibfnamefont {K.}~\bibnamefont
  {Yagi}}, \bibinfo {author} {\bibfnamefont {T.}~\bibnamefont {Hatsuda}}, \
  and\ \bibinfo {author} {\bibfnamefont {Y.}~\bibnamefont {Miake}},\
  }\href@noop {} {\emph {\bibinfo {title} {{Quark-gluon plasma: From big bang
  to little bang}}}},\ Vol.~\bibinfo {volume} {23}\ (\bibinfo {year}
  {2005})\BibitemShut {NoStop}%
\bibitem [{\citenamefont {Skokov}\ \emph {et~al.}(2009)\citenamefont {Skokov},
  \citenamefont {Illarionov},\ and\ \citenamefont {Toneev}}]{Skokov:2009qp}%
  \BibitemOpen
  \bibfield  {author} {\bibinfo {author} {\bibfnamefont {V.}~\bibnamefont
  {Skokov}}, \bibinfo {author} {\bibfnamefont {A.~Y.}\ \bibnamefont
  {Illarionov}}, \ and\ \bibinfo {author} {\bibfnamefont {V.}~\bibnamefont
  {Toneev}},\ }\href {\doibase 10.1142/S0217751X09047570} {\bibfield  {journal}
  {\bibinfo  {journal} {Int. J. Mod. Phys. A}\ }\textbf {\bibinfo {volume}
  {24}},\ \bibinfo {pages} {5925} (\bibinfo {year} {2009})},\ \Eprint
  {http://arxiv.org/abs/0907.1396} {arXiv:0907.1396 [nucl-th]} \BibitemShut
  {NoStop}%
\bibitem [{\citenamefont {Kharzeev}\ \emph {et~al.}(2013)\citenamefont
  {Kharzeev}, \citenamefont {Landsteiner}, \citenamefont {Schmitt},\ and\
  \citenamefont {Yee}}]{Kharzeev:2013jha}%
  \BibitemOpen
  \bibinfo {editor} {\bibfnamefont {D.}~\bibnamefont {Kharzeev}}, \bibinfo
  {editor} {\bibfnamefont {K.}~\bibnamefont {Landsteiner}}, \bibinfo {editor}
  {\bibfnamefont {A.}~\bibnamefont {Schmitt}}, \ and\ \bibinfo {editor}
  {\bibfnamefont {H.-U.}\ \bibnamefont {Yee}},\ eds.,\ \href {\doibase
  10.1007/978-3-642-37305-3} {\emph {\bibinfo {title} {{Strongly Interacting
  Matter in Magnetic Fields}}}},\ Vol.\ \bibinfo {volume} {871}\ (\bibinfo
  {year} {2013})\BibitemShut {NoStop}%
\bibitem [{\citenamefont {Miransky}\ and\ \citenamefont
  {Shovkovy}(2015)}]{Miransky:2015ava}%
  \BibitemOpen
  \bibfield  {author} {\bibinfo {author} {\bibfnamefont {V.~A.}\ \bibnamefont
  {Miransky}}\ and\ \bibinfo {author} {\bibfnamefont {I.~A.}\ \bibnamefont
  {Shovkovy}},\ }\href {\doibase 10.1016/j.physrep.2015.02.003} {\bibfield
  {journal} {\bibinfo  {journal} {Phys. Rept.}\ }\textbf {\bibinfo {volume}
  {576}},\ \bibinfo {pages} {1} (\bibinfo {year} {2015})},\ \Eprint
  {http://arxiv.org/abs/1503.00732} {arXiv:1503.00732 [hep-ph]} \BibitemShut
  {NoStop}%
\bibitem [{\citenamefont {Gusynin}\ \emph {et~al.}(1994)\citenamefont
  {Gusynin}, \citenamefont {Miransky},\ and\ \citenamefont
  {Shovkovy}}]{Gusynin:1994re}%
  \BibitemOpen
  \bibfield  {author} {\bibinfo {author} {\bibfnamefont {V.~P.}\ \bibnamefont
  {Gusynin}}, \bibinfo {author} {\bibfnamefont {V.~A.}\ \bibnamefont
  {Miransky}}, \ and\ \bibinfo {author} {\bibfnamefont {I.~A.}\ \bibnamefont
  {Shovkovy}},\ }\href {\doibase 10.1103/PhysRevLett.73.3499} {\bibfield
  {journal} {\bibinfo  {journal} {Phys. Rev. Lett.}\ }\textbf {\bibinfo
  {volume} {73}},\ \bibinfo {pages} {3499} (\bibinfo {year} {1994})},\ \bibinfo
  {note} {[Erratum: Phys.Rev.Lett. 76, 1005 (1996)]},\ \Eprint
  {http://arxiv.org/abs/hep-ph/9405262} {arXiv:hep-ph/9405262} \BibitemShut
  {NoStop}%
\bibitem [{\citenamefont {Bali}\ \emph
  {et~al.}(2012{\natexlab{a}})\citenamefont {Bali}, \citenamefont {Bruckmann},
  \citenamefont {Endrodi}, \citenamefont {Fodor}, \citenamefont {Katz},\ and\
  \citenamefont {Schafer}}]{Bali:2012zg}%
  \BibitemOpen
  \bibfield  {author} {\bibinfo {author} {\bibfnamefont {G.~S.}\ \bibnamefont
  {Bali}}, \bibinfo {author} {\bibfnamefont {F.}~\bibnamefont {Bruckmann}},
  \bibinfo {author} {\bibfnamefont {G.}~\bibnamefont {Endrodi}}, \bibinfo
  {author} {\bibfnamefont {Z.}~\bibnamefont {Fodor}}, \bibinfo {author}
  {\bibfnamefont {S.~D.}\ \bibnamefont {Katz}}, \ and\ \bibinfo {author}
  {\bibfnamefont {A.}~\bibnamefont {Schafer}},\ }\href {\doibase
  10.1103/PhysRevD.86.071502} {\bibfield  {journal} {\bibinfo  {journal} {Phys.
  Rev. D}\ }\textbf {\bibinfo {volume} {86}},\ \bibinfo {pages} {071502}
  (\bibinfo {year} {2012}{\natexlab{a}})},\ \Eprint
  {http://arxiv.org/abs/1206.4205} {arXiv:1206.4205 [hep-lat]} \BibitemShut
  {NoStop}%
\bibitem [{\citenamefont {Bali}\ \emph
  {et~al.}(2012{\natexlab{b}})\citenamefont {Bali}, \citenamefont {Bruckmann},
  \citenamefont {Endrodi}, \citenamefont {Fodor}, \citenamefont {Katz},
  \citenamefont {Krieg}, \citenamefont {Schafer},\ and\ \citenamefont
  {Szabo}}]{Bali:2011qj}%
  \BibitemOpen
  \bibfield  {author} {\bibinfo {author} {\bibfnamefont {G.~S.}\ \bibnamefont
  {Bali}}, \bibinfo {author} {\bibfnamefont {F.}~\bibnamefont {Bruckmann}},
  \bibinfo {author} {\bibfnamefont {G.}~\bibnamefont {Endrodi}}, \bibinfo
  {author} {\bibfnamefont {Z.}~\bibnamefont {Fodor}}, \bibinfo {author}
  {\bibfnamefont {S.~D.}\ \bibnamefont {Katz}}, \bibinfo {author}
  {\bibfnamefont {S.}~\bibnamefont {Krieg}}, \bibinfo {author} {\bibfnamefont
  {A.}~\bibnamefont {Schafer}}, \ and\ \bibinfo {author} {\bibfnamefont
  {K.~K.}\ \bibnamefont {Szabo}},\ }\href {\doibase 10.1007/JHEP02(2012)044}
  {\bibfield  {journal} {\bibinfo  {journal} {JHEP}\ }\textbf {\bibinfo
  {volume} {02}},\ \bibinfo {pages} {044} (\bibinfo {year}
  {2012}{\natexlab{b}})},\ \Eprint {http://arxiv.org/abs/1111.4956}
  {arXiv:1111.4956 [hep-lat]} \BibitemShut {NoStop}%
\bibitem [{\citenamefont {Fukushima}\ \emph {et~al.}(2008)\citenamefont
  {Fukushima}, \citenamefont {Kharzeev},\ and\ \citenamefont
  {Warringa}}]{Fukushima:2008xe}%
  \BibitemOpen
  \bibfield  {author} {\bibinfo {author} {\bibfnamefont {K.}~\bibnamefont
  {Fukushima}}, \bibinfo {author} {\bibfnamefont {D.~E.}\ \bibnamefont
  {Kharzeev}}, \ and\ \bibinfo {author} {\bibfnamefont {H.~J.}\ \bibnamefont
  {Warringa}},\ }\href {\doibase 10.1103/PhysRevD.78.074033} {\bibfield
  {journal} {\bibinfo  {journal} {Phys. Rev. D}\ }\textbf {\bibinfo {volume}
  {78}},\ \bibinfo {pages} {074033} (\bibinfo {year} {2008})},\ \Eprint
  {http://arxiv.org/abs/0808.3382} {arXiv:0808.3382 [hep-ph]} \BibitemShut
  {NoStop}%
\bibitem [{\citenamefont {Tuchin}(2016)}]{Tuchin:2015oka}%
  \BibitemOpen
  \bibfield  {author} {\bibinfo {author} {\bibfnamefont {K.}~\bibnamefont
  {Tuchin}},\ }\href {\doibase 10.1103/PhysRevC.93.014905} {\bibfield
  {journal} {\bibinfo  {journal} {Phys. Rev. C}\ }\textbf {\bibinfo {volume}
  {93}},\ \bibinfo {pages} {014905} (\bibinfo {year} {2016})},\ \Eprint
  {http://arxiv.org/abs/1508.06925} {arXiv:1508.06925 [hep-ph]} \BibitemShut
  {NoStop}%
\bibitem [{\citenamefont {Guo}\ \emph {et~al.}(2020)\citenamefont {Guo},
  \citenamefont {Liao},\ and\ \citenamefont {Wang}}]{Guo:2019mgh}%
  \BibitemOpen
  \bibfield  {author} {\bibinfo {author} {\bibfnamefont {X.}~\bibnamefont
  {Guo}}, \bibinfo {author} {\bibfnamefont {J.}~\bibnamefont {Liao}}, \ and\
  \bibinfo {author} {\bibfnamefont {E.}~\bibnamefont {Wang}},\ }\href {\doibase
  10.1038/s41598-020-59129-6} {\bibfield  {journal} {\bibinfo  {journal} {Sci.
  Rep.}\ }\textbf {\bibinfo {volume} {10}},\ \bibinfo {pages} {2196} (\bibinfo
  {year} {2020})},\ \Eprint {http://arxiv.org/abs/1904.04704} {arXiv:1904.04704
  [hep-ph]} \BibitemShut {NoStop}%
\bibitem [{\citenamefont {Wang}\ \emph {et~al.}(2022)\citenamefont {Wang},
  \citenamefont {Zhao}, \citenamefont {Greiner}, \citenamefont {Xu},\ and\
  \citenamefont {Zhuang}}]{Wang:2021oqq}%
  \BibitemOpen
  \bibfield  {author} {\bibinfo {author} {\bibfnamefont {Z.}~\bibnamefont
  {Wang}}, \bibinfo {author} {\bibfnamefont {J.}~\bibnamefont {Zhao}}, \bibinfo
  {author} {\bibfnamefont {C.}~\bibnamefont {Greiner}}, \bibinfo {author}
  {\bibfnamefont {Z.}~\bibnamefont {Xu}}, \ and\ \bibinfo {author}
  {\bibfnamefont {P.}~\bibnamefont {Zhuang}},\ }\href {\doibase
  10.1103/PhysRevC.105.L041901} {\bibfield  {journal} {\bibinfo  {journal}
  {Phys. Rev. C}\ }\textbf {\bibinfo {volume} {105}},\ \bibinfo {pages}
  {L041901} (\bibinfo {year} {2022})},\ \Eprint
  {http://arxiv.org/abs/2110.14302} {arXiv:2110.14302 [hep-ph]} \BibitemShut
  {NoStop}%
\bibitem [{\citenamefont {Abdallah}\ \emph {et~al.}(2022)\citenamefont
  {Abdallah} \emph {et~al.}}]{STAR:2021mii}%
  \BibitemOpen
  \bibfield  {author} {\bibinfo {author} {\bibfnamefont {M.}~\bibnamefont
  {Abdallah}} \emph {et~al.} (\bibinfo {collaboration} {STAR}),\ }\href
  {\doibase 10.1103/PhysRevC.105.014901} {\bibfield  {journal} {\bibinfo
  {journal} {Phys. Rev. C}\ }\textbf {\bibinfo {volume} {105}},\ \bibinfo
  {pages} {014901} (\bibinfo {year} {2022})},\ \Eprint
  {http://arxiv.org/abs/2109.00131} {arXiv:2109.00131 [nucl-ex]} \BibitemShut
  {NoStop}%
\bibitem [{\citenamefont {Forster}(2018)}]{Forster:1975pm}%
  \BibitemOpen
  \bibfield  {author} {\bibinfo {author} {\bibfnamefont {D.}~\bibnamefont
  {Forster}},\ }\href {\doibase https://doi.org/10.1201/9780429493683} {\emph
  {\bibinfo {title} {{Hydrodynamics Fluctuation, Broken Symmetry and
  Correlation Function}}}}\ (\bibinfo  {publisher} {CRC Press},\ \bibinfo
  {year} {2018})\BibitemShut {NoStop}%
\bibitem [{\citenamefont {Callen}\ and\ \citenamefont
  {Welton}(1951)}]{Callen:1951vq}%
  \BibitemOpen
  \bibfield  {author} {\bibinfo {author} {\bibfnamefont {H.~B.}\ \bibnamefont
  {Callen}}\ and\ \bibinfo {author} {\bibfnamefont {T.~A.}\ \bibnamefont
  {Welton}},\ }\href {\doibase 10.1103/PhysRev.83.34} {\bibfield  {journal}
  {\bibinfo  {journal} {Phys. Rev.}\ }\textbf {\bibinfo {volume} {83}},\
  \bibinfo {pages} {34} (\bibinfo {year} {1951})}\BibitemShut {NoStop}%
\bibitem [{\citenamefont {Kubo}(1957)}]{Kubo:1957mj}%
  \BibitemOpen
  \bibfield  {author} {\bibinfo {author} {\bibfnamefont {R.}~\bibnamefont
  {Kubo}},\ }\href {\doibase 10.1143/JPSJ.12.570} {\bibfield  {journal}
  {\bibinfo  {journal} {J. Phys. Soc. Jap.}\ }\textbf {\bibinfo {volume}
  {12}},\ \bibinfo {pages} {570} (\bibinfo {year} {1957})}\BibitemShut
  {NoStop}%
\bibitem [{\citenamefont {Weldon}(1990)}]{Weldon:1990iw}%
  \BibitemOpen
  \bibfield  {author} {\bibinfo {author} {\bibfnamefont {H.~A.}\ \bibnamefont
  {Weldon}},\ }\href {\doibase 10.1103/PhysRevD.42.2384} {\bibfield  {journal}
  {\bibinfo  {journal} {Phys. Rev. D}\ }\textbf {\bibinfo {volume} {42}},\
  \bibinfo {pages} {2384} (\bibinfo {year} {1990})}\BibitemShut {NoStop}%
\bibitem [{\citenamefont {Islam}\ \emph {et~al.}(2015)\citenamefont {Islam},
  \citenamefont {Majumder}, \citenamefont {Haque},\ and\ \citenamefont
  {Mustafa}}]{Islam:2014sea}%
  \BibitemOpen
  \bibfield  {author} {\bibinfo {author} {\bibfnamefont {C.~A.}\ \bibnamefont
  {Islam}}, \bibinfo {author} {\bibfnamefont {S.}~\bibnamefont {Majumder}},
  \bibinfo {author} {\bibfnamefont {N.}~\bibnamefont {Haque}}, \ and\ \bibinfo
  {author} {\bibfnamefont {M.~G.}\ \bibnamefont {Mustafa}},\ }\href {\doibase
  10.1007/JHEP02(2015)011} {\bibfield  {journal} {\bibinfo  {journal} {JHEP}\
  }\textbf {\bibinfo {volume} {02}},\ \bibinfo {pages} {011} (\bibinfo {year}
  {2015})},\ \Eprint {http://arxiv.org/abs/1411.6407} {arXiv:1411.6407
  [hep-ph]} \BibitemShut {NoStop}%
\bibitem [{\citenamefont {Gale}\ \emph {et~al.}(2015)\citenamefont {Gale},
  \citenamefont {Hidaka}, \citenamefont {Jeon}, \citenamefont {Lin},
  \citenamefont {Paquet}, \citenamefont {Pisarski}, \citenamefont {Satow},
  \citenamefont {Skokov},\ and\ \citenamefont {Vujanovic}}]{Gale:2014dfa}%
  \BibitemOpen
  \bibfield  {author} {\bibinfo {author} {\bibfnamefont {C.}~\bibnamefont
  {Gale}}, \bibinfo {author} {\bibfnamefont {Y.}~\bibnamefont {Hidaka}},
  \bibinfo {author} {\bibfnamefont {S.}~\bibnamefont {Jeon}}, \bibinfo {author}
  {\bibfnamefont {S.}~\bibnamefont {Lin}}, \bibinfo {author} {\bibfnamefont
  {J.-F.}\ \bibnamefont {Paquet}}, \bibinfo {author} {\bibfnamefont {R.~D.}\
  \bibnamefont {Pisarski}}, \bibinfo {author} {\bibfnamefont {D.}~\bibnamefont
  {Satow}}, \bibinfo {author} {\bibfnamefont {V.~V.}\ \bibnamefont {Skokov}}, \
  and\ \bibinfo {author} {\bibfnamefont {G.}~\bibnamefont {Vujanovic}},\ }\href
  {\doibase 10.1103/PhysRevLett.114.072301} {\bibfield  {journal} {\bibinfo
  {journal} {Phys. Rev. Lett.}\ }\textbf {\bibinfo {volume} {114}},\ \bibinfo
  {pages} {072301} (\bibinfo {year} {2015})},\ \Eprint
  {http://arxiv.org/abs/1409.4778} {arXiv:1409.4778 [hep-ph]} \BibitemShut
  {NoStop}%
\bibitem [{\citenamefont {Hidaka}\ \emph {et~al.}(2015)\citenamefont {Hidaka},
  \citenamefont {Lin}, \citenamefont {Pisarski},\ and\ \citenamefont
  {Satow}}]{Hidaka:2015ima}%
  \BibitemOpen
  \bibfield  {author} {\bibinfo {author} {\bibfnamefont {Y.}~\bibnamefont
  {Hidaka}}, \bibinfo {author} {\bibfnamefont {S.}~\bibnamefont {Lin}},
  \bibinfo {author} {\bibfnamefont {R.~D.}\ \bibnamefont {Pisarski}}, \ and\
  \bibinfo {author} {\bibfnamefont {D.}~\bibnamefont {Satow}},\ }\href
  {\doibase 10.1007/JHEP10(2015)005} {\bibfield  {journal} {\bibinfo  {journal}
  {JHEP}\ }\textbf {\bibinfo {volume} {10}},\ \bibinfo {pages} {005} (\bibinfo
  {year} {2015})},\ \Eprint {http://arxiv.org/abs/1504.01770} {arXiv:1504.01770
  [hep-ph]} \BibitemShut {NoStop}%
\bibitem [{\citenamefont {Tuchin}(2013{\natexlab{a}})}]{Tuchin:2012mf}%
  \BibitemOpen
  \bibfield  {author} {\bibinfo {author} {\bibfnamefont {K.}~\bibnamefont
  {Tuchin}},\ }\href {\doibase 10.1103/PhysRevC.87.024912} {\bibfield
  {journal} {\bibinfo  {journal} {Phys. Rev. C}\ }\textbf {\bibinfo {volume}
  {87}},\ \bibinfo {pages} {024912} (\bibinfo {year} {2013}{\natexlab{a}})},\
  \Eprint {http://arxiv.org/abs/1206.0485} {arXiv:1206.0485 [hep-ph]}
  \BibitemShut {NoStop}%
\bibitem [{\citenamefont {Tuchin}(2013{\natexlab{b}})}]{Tuchin:2013bda}%
  \BibitemOpen
  \bibfield  {author} {\bibinfo {author} {\bibfnamefont {K.}~\bibnamefont
  {Tuchin}},\ }\href {\doibase 10.1103/PhysRevC.88.024910} {\bibfield
  {journal} {\bibinfo  {journal} {Phys. Rev. C}\ }\textbf {\bibinfo {volume}
  {88}},\ \bibinfo {pages} {024910} (\bibinfo {year} {2013}{\natexlab{b}})},\
  \Eprint {http://arxiv.org/abs/1305.0545} {arXiv:1305.0545 [nucl-th]}
  \BibitemShut {NoStop}%
\bibitem [{\citenamefont {Tuchin}(2013{\natexlab{c}})}]{Tuchin:2013ie}%
  \BibitemOpen
  \bibfield  {author} {\bibinfo {author} {\bibfnamefont {K.}~\bibnamefont
  {Tuchin}},\ }\href {\doibase 10.1155/2013/490495} {\bibfield  {journal}
  {\bibinfo  {journal} {Adv. High Energy Phys.}\ }\textbf {\bibinfo {volume}
  {2013}},\ \bibinfo {pages} {490495} (\bibinfo {year} {2013}{\natexlab{c}})},\
  \Eprint {http://arxiv.org/abs/1301.0099} {arXiv:1301.0099 [hep-ph]}
  \BibitemShut {NoStop}%
\bibitem [{\citenamefont {Sadooghi}\ and\ \citenamefont
  {Taghinavaz}(2017)}]{Sadooghi:2016jyf}%
  \BibitemOpen
  \bibfield  {author} {\bibinfo {author} {\bibfnamefont {N.}~\bibnamefont
  {Sadooghi}}\ and\ \bibinfo {author} {\bibfnamefont {F.}~\bibnamefont
  {Taghinavaz}},\ }\href {\doibase 10.1016/j.aop.2016.11.008} {\bibfield
  {journal} {\bibinfo  {journal} {Annals Phys.}\ }\textbf {\bibinfo {volume}
  {376}},\ \bibinfo {pages} {218} (\bibinfo {year} {2017})},\ \Eprint
  {http://arxiv.org/abs/1601.04887} {arXiv:1601.04887 [hep-ph]} \BibitemShut
  {NoStop}%
\bibitem [{\citenamefont {Hattori}\ \emph {et~al.}(2021)\citenamefont
  {Hattori}, \citenamefont {Taya},\ and\ \citenamefont
  {Yoshida}}]{Hattori:2020htm}%
  \BibitemOpen
  \bibfield  {author} {\bibinfo {author} {\bibfnamefont {K.}~\bibnamefont
  {Hattori}}, \bibinfo {author} {\bibfnamefont {H.}~\bibnamefont {Taya}}, \
  and\ \bibinfo {author} {\bibfnamefont {S.}~\bibnamefont {Yoshida}},\ }\href
  {\doibase 10.1007/JHEP01(2021)093} {\bibfield  {journal} {\bibinfo  {journal}
  {JHEP}\ }\textbf {\bibinfo {volume} {01}},\ \bibinfo {pages} {093} (\bibinfo
  {year} {2021})},\ \Eprint {http://arxiv.org/abs/2010.13492} {arXiv:2010.13492
  [hep-ph]} \BibitemShut {NoStop}%
\bibitem [{\citenamefont {Bandyopadhyay}\ \emph {et~al.}(2016)\citenamefont
  {Bandyopadhyay}, \citenamefont {Islam},\ and\ \citenamefont
  {Mustafa}}]{Bandyopadhyay:2016fyd}%
  \BibitemOpen
  \bibfield  {author} {\bibinfo {author} {\bibfnamefont {A.}~\bibnamefont
  {Bandyopadhyay}}, \bibinfo {author} {\bibfnamefont {C.~A.}\ \bibnamefont
  {Islam}}, \ and\ \bibinfo {author} {\bibfnamefont {M.~G.}\ \bibnamefont
  {Mustafa}},\ }\href {\doibase 10.1103/PhysRevD.94.114034} {\bibfield
  {journal} {\bibinfo  {journal} {Phys. Rev. D}\ }\textbf {\bibinfo {volume}
  {94}},\ \bibinfo {pages} {114034} (\bibinfo {year} {2016})},\ \Eprint
  {http://arxiv.org/abs/1602.06769} {arXiv:1602.06769 [hep-ph]} \BibitemShut
  {NoStop}%
\bibitem [{\citenamefont {Bandyopadhyay}\ and\ \citenamefont
  {Mallik}(2017)}]{Bandyopadhyay:2017raf}%
  \BibitemOpen
  \bibfield  {author} {\bibinfo {author} {\bibfnamefont {A.}~\bibnamefont
  {Bandyopadhyay}}\ and\ \bibinfo {author} {\bibfnamefont {S.}~\bibnamefont
  {Mallik}},\ }\href {\doibase 10.1103/PhysRevD.95.074019} {\bibfield
  {journal} {\bibinfo  {journal} {Phys. Rev. D}\ }\textbf {\bibinfo {volume}
  {95}},\ \bibinfo {pages} {074019} (\bibinfo {year} {2017})},\ \Eprint
  {http://arxiv.org/abs/1704.01364} {arXiv:1704.01364 [hep-ph]} \BibitemShut
  {NoStop}%
\bibitem [{\citenamefont {Ghosh}\ and\ \citenamefont
  {Chandra}(2018)}]{Ghosh:2018xhh}%
  \BibitemOpen
  \bibfield  {author} {\bibinfo {author} {\bibfnamefont {S.}~\bibnamefont
  {Ghosh}}\ and\ \bibinfo {author} {\bibfnamefont {V.}~\bibnamefont
  {Chandra}},\ }\href {\doibase 10.1103/PhysRevD.98.076006} {\bibfield
  {journal} {\bibinfo  {journal} {Phys. Rev. D}\ }\textbf {\bibinfo {volume}
  {98}},\ \bibinfo {pages} {076006} (\bibinfo {year} {2018})},\ \Eprint
  {http://arxiv.org/abs/1808.05176} {arXiv:1808.05176 [hep-ph]} \BibitemShut
  {NoStop}%
\bibitem [{\citenamefont {Islam}\ \emph {et~al.}(2019)\citenamefont {Islam},
  \citenamefont {Bandyopadhyay}, \citenamefont {Roy},\ and\ \citenamefont
  {Sarkar}}]{Islam:2018sog}%
  \BibitemOpen
  \bibfield  {author} {\bibinfo {author} {\bibfnamefont {C.~A.}\ \bibnamefont
  {Islam}}, \bibinfo {author} {\bibfnamefont {A.}~\bibnamefont
  {Bandyopadhyay}}, \bibinfo {author} {\bibfnamefont {P.~K.}\ \bibnamefont
  {Roy}}, \ and\ \bibinfo {author} {\bibfnamefont {S.}~\bibnamefont {Sarkar}},\
  }\href {\doibase 10.1103/PhysRevD.99.094028} {\bibfield  {journal} {\bibinfo
  {journal} {Phys. Rev. D}\ }\textbf {\bibinfo {volume} {99}},\ \bibinfo
  {pages} {094028} (\bibinfo {year} {2019})},\ \Eprint
  {http://arxiv.org/abs/1812.10380} {arXiv:1812.10380 [hep-ph]} \BibitemShut
  {NoStop}%
\bibitem [{\citenamefont {Ghosh}\ \emph {et~al.}(2020)\citenamefont {Ghosh},
  \citenamefont {Chaudhuri}, \citenamefont {Sarkar},\ and\ \citenamefont
  {Roy}}]{Ghosh:2020xwp}%
  \BibitemOpen
  \bibfield  {author} {\bibinfo {author} {\bibfnamefont {S.}~\bibnamefont
  {Ghosh}}, \bibinfo {author} {\bibfnamefont {N.}~\bibnamefont {Chaudhuri}},
  \bibinfo {author} {\bibfnamefont {S.}~\bibnamefont {Sarkar}}, \ and\ \bibinfo
  {author} {\bibfnamefont {P.}~\bibnamefont {Roy}},\ }\href {\doibase
  10.1103/PhysRevD.101.096002} {\bibfield  {journal} {\bibinfo  {journal}
  {Phys. Rev. D}\ }\textbf {\bibinfo {volume} {101}},\ \bibinfo {pages}
  {096002} (\bibinfo {year} {2020})},\ \Eprint
  {http://arxiv.org/abs/2004.09203} {arXiv:2004.09203 [nucl-th]} \BibitemShut
  {NoStop}%
\bibitem [{\citenamefont {Wang}\ \emph {et~al.}(2020)\citenamefont {Wang},
  \citenamefont {Shovkovy}, \citenamefont {Yu},\ and\ \citenamefont
  {Huang}}]{Wang:2020dsr}%
  \BibitemOpen
  \bibfield  {author} {\bibinfo {author} {\bibfnamefont {X.}~\bibnamefont
  {Wang}}, \bibinfo {author} {\bibfnamefont {I.~A.}\ \bibnamefont {Shovkovy}},
  \bibinfo {author} {\bibfnamefont {L.}~\bibnamefont {Yu}}, \ and\ \bibinfo
  {author} {\bibfnamefont {M.}~\bibnamefont {Huang}},\ }\href {\doibase
  10.1103/PhysRevD.102.076010} {\bibfield  {journal} {\bibinfo  {journal}
  {Phys. Rev. D}\ }\textbf {\bibinfo {volume} {102}},\ \bibinfo {pages}
  {076010} (\bibinfo {year} {2020})},\ \Eprint
  {http://arxiv.org/abs/2006.16254} {arXiv:2006.16254 [hep-ph]} \BibitemShut
  {NoStop}%
\bibitem [{\citenamefont {Wang}\ and\ \citenamefont
  {Shovkovy}(2021)}]{Wang:2021ebh}%
  \BibitemOpen
  \bibfield  {author} {\bibinfo {author} {\bibfnamefont {X.}~\bibnamefont
  {Wang}}\ and\ \bibinfo {author} {\bibfnamefont {I.}~\bibnamefont
  {Shovkovy}},\ }\href@noop {} {\  (\bibinfo {year} {2021})},\ \Eprint
  {http://arxiv.org/abs/2103.01967} {arXiv:2103.01967 [nucl-th]} \BibitemShut
  {NoStop}%
\bibitem [{\citenamefont {Schwinger}(1951)}]{Schwinger:1951nm}%
  \BibitemOpen
  \bibfield  {author} {\bibinfo {author} {\bibfnamefont {J.~S.}\ \bibnamefont
  {Schwinger}},\ }\href {\doibase 10.1103/PhysRev.82.664} {\bibfield  {journal}
  {\bibinfo  {journal} {Phys. Rev.}\ }\textbf {\bibinfo {volume} {82}},\
  \bibinfo {pages} {664} (\bibinfo {year} {1951})}\BibitemShut {NoStop}%
\bibitem [{\citenamefont {Wang}\ and\ \citenamefont
  {Shovkovy}(2022)}]{Wang:2022jxx}%
  \BibitemOpen
  \bibfield  {author} {\bibinfo {author} {\bibfnamefont {X.}~\bibnamefont
  {Wang}}\ and\ \bibinfo {author} {\bibfnamefont {I.~A.}\ \bibnamefont
  {Shovkovy}},\ }\href {\doibase 10.1103/PhysRevD.106.036014} {\bibfield
  {journal} {\bibinfo  {journal} {Phys. Rev. D}\ }\textbf {\bibinfo {volume}
  {106}},\ \bibinfo {pages} {036014} (\bibinfo {year} {2022})},\ \Eprint
  {http://arxiv.org/abs/2205.00276} {arXiv:2205.00276 [nucl-th]} \BibitemShut
  {NoStop}%
\bibitem [{\citenamefont {Farias}\ \emph {et~al.}(2014)\citenamefont {Farias},
  \citenamefont {Gomes}, \citenamefont {Krein},\ and\ \citenamefont
  {Pinto}}]{Farias:2014eca}%
  \BibitemOpen
  \bibfield  {author} {\bibinfo {author} {\bibfnamefont {R.~L.~S.}\
  \bibnamefont {Farias}}, \bibinfo {author} {\bibfnamefont {K.~P.}\
  \bibnamefont {Gomes}}, \bibinfo {author} {\bibfnamefont {G.~I.}\ \bibnamefont
  {Krein}}, \ and\ \bibinfo {author} {\bibfnamefont {M.~B.}\ \bibnamefont
  {Pinto}},\ }\href {\doibase 10.1103/PhysRevC.90.025203} {\bibfield  {journal}
  {\bibinfo  {journal} {Phys. Rev. C}\ }\textbf {\bibinfo {volume} {90}},\
  \bibinfo {pages} {025203} (\bibinfo {year} {2014})},\ \Eprint
  {http://arxiv.org/abs/1404.3931} {arXiv:1404.3931 [hep-ph]} \BibitemShut
  {NoStop}%
\bibitem [{\citenamefont {Farias}\ \emph {et~al.}(2017)\citenamefont {Farias},
  \citenamefont {Timoteo}, \citenamefont {Avancini}, \citenamefont {Pinto},\
  and\ \citenamefont {Krein}}]{Farias:2016gmy}%
  \BibitemOpen
  \bibfield  {author} {\bibinfo {author} {\bibfnamefont {R.~L.~S.}\
  \bibnamefont {Farias}}, \bibinfo {author} {\bibfnamefont {V.~S.}\
  \bibnamefont {Timoteo}}, \bibinfo {author} {\bibfnamefont {S.~S.}\
  \bibnamefont {Avancini}}, \bibinfo {author} {\bibfnamefont {M.~B.}\
  \bibnamefont {Pinto}}, \ and\ \bibinfo {author} {\bibfnamefont
  {G.}~\bibnamefont {Krein}},\ }\href {\doibase 10.1140/epja/i2017-12320-8}
  {\bibfield  {journal} {\bibinfo  {journal} {Eur. Phys. J. A}\ }\textbf
  {\bibinfo {volume} {53}},\ \bibinfo {pages} {101} (\bibinfo {year} {2017})},\
  \Eprint {http://arxiv.org/abs/1603.03847} {arXiv:1603.03847 [hep-ph]}
  \BibitemShut {NoStop}%
\bibitem [{\citenamefont {Cleymans}\ \emph {et~al.}(1987)\citenamefont
  {Cleymans}, \citenamefont {Fingberg},\ and\ \citenamefont
  {Redlich}}]{Cleymans:1986na}%
  \BibitemOpen
  \bibfield  {author} {\bibinfo {author} {\bibfnamefont {J.}~\bibnamefont
  {Cleymans}}, \bibinfo {author} {\bibfnamefont {J.}~\bibnamefont {Fingberg}},
  \ and\ \bibinfo {author} {\bibfnamefont {K.}~\bibnamefont {Redlich}},\ }\href
  {\doibase 10.1103/PhysRevD.35.2153} {\bibfield  {journal} {\bibinfo
  {journal} {Phys. Rev. D}\ }\textbf {\bibinfo {volume} {35}},\ \bibinfo
  {pages} {2153} (\bibinfo {year} {1987})}\BibitemShut {NoStop}%
\bibitem [{\citenamefont {Greiner}\ \emph {et~al.}(2011)\citenamefont
  {Greiner}, \citenamefont {Haque}, \citenamefont {Mustafa},\ and\
  \citenamefont {Thoma}}]{Greiner:2010zg}%
  \BibitemOpen
  \bibfield  {author} {\bibinfo {author} {\bibfnamefont {C.}~\bibnamefont
  {Greiner}}, \bibinfo {author} {\bibfnamefont {N.}~\bibnamefont {Haque}},
  \bibinfo {author} {\bibfnamefont {M.~G.}\ \bibnamefont {Mustafa}}, \ and\
  \bibinfo {author} {\bibfnamefont {M.~H.}\ \bibnamefont {Thoma}},\ }\href
  {\doibase 10.1103/PhysRevC.83.014908} {\bibfield  {journal} {\bibinfo
  {journal} {Phys. Rev. C}\ }\textbf {\bibinfo {volume} {83}},\ \bibinfo
  {pages} {014908} (\bibinfo {year} {2011})},\ \Eprint
  {http://arxiv.org/abs/1010.2169} {arXiv:1010.2169 [hep-ph]} \BibitemShut
  {NoStop}%
\bibitem [{\citenamefont {Gusynin}\ \emph {et~al.}(1996)\citenamefont
  {Gusynin}, \citenamefont {Miransky},\ and\ \citenamefont
  {Shovkovy}}]{Gusynin:1995nb}%
  \BibitemOpen
  \bibfield  {author} {\bibinfo {author} {\bibfnamefont {V.~P.}\ \bibnamefont
  {Gusynin}}, \bibinfo {author} {\bibfnamefont {V.~A.}\ \bibnamefont
  {Miransky}}, \ and\ \bibinfo {author} {\bibfnamefont {I.~A.}\ \bibnamefont
  {Shovkovy}},\ }\href {\doibase 10.1016/0550-3213(96)00021-1} {\bibfield
  {journal} {\bibinfo  {journal} {Nucl. Phys. B}\ }\textbf {\bibinfo {volume}
  {462}},\ \bibinfo {pages} {249} (\bibinfo {year} {1996})},\ \Eprint
  {http://arxiv.org/abs/hep-ph/9509320} {arXiv:hep-ph/9509320} \BibitemShut
  {NoStop}%
\bibitem [{\citenamefont {Nambu}\ and\ \citenamefont
  {Jona-Lasinio}(1961{\natexlab{a}})}]{Nambu:1961tp}%
  \BibitemOpen
  \bibfield  {author} {\bibinfo {author} {\bibfnamefont {Y.}~\bibnamefont
  {Nambu}}\ and\ \bibinfo {author} {\bibfnamefont {G.}~\bibnamefont
  {Jona-Lasinio}},\ }\href {\doibase 10.1103/PhysRev.122.345} {\bibfield
  {journal} {\bibinfo  {journal} {Phys. Rev.}\ }\textbf {\bibinfo {volume}
  {122}},\ \bibinfo {pages} {345} (\bibinfo {year}
  {1961}{\natexlab{a}})}\BibitemShut {NoStop}%
\bibitem [{\citenamefont {Nambu}\ and\ \citenamefont
  {Jona-Lasinio}(1961{\natexlab{b}})}]{Nambu:1961fr}%
  \BibitemOpen
  \bibfield  {author} {\bibinfo {author} {\bibfnamefont {Y.}~\bibnamefont
  {Nambu}}\ and\ \bibinfo {author} {\bibfnamefont {G.}~\bibnamefont
  {Jona-Lasinio}},\ }\href {\doibase 10.1103/PhysRev.124.246} {\bibfield
  {journal} {\bibinfo  {journal} {Phys. Rev.}\ }\textbf {\bibinfo {volume}
  {124}},\ \bibinfo {pages} {246} (\bibinfo {year}
  {1961}{\natexlab{b}})}\BibitemShut {NoStop}%
\bibitem [{\citenamefont {Chaudhuri}\ \emph {et~al.}(2021)\citenamefont
  {Chaudhuri}, \citenamefont {Ghosh}, \citenamefont {Sarkar},\ and\
  \citenamefont {Roy}}]{Chaudhuri:2021skc}%
  \BibitemOpen
  \bibfield  {author} {\bibinfo {author} {\bibfnamefont {N.}~\bibnamefont
  {Chaudhuri}}, \bibinfo {author} {\bibfnamefont {S.}~\bibnamefont {Ghosh}},
  \bibinfo {author} {\bibfnamefont {S.}~\bibnamefont {Sarkar}}, \ and\ \bibinfo
  {author} {\bibfnamefont {P.}~\bibnamefont {Roy}},\ }\href {\doibase
  10.1103/PhysRevD.103.096021} {\bibfield  {journal} {\bibinfo  {journal}
  {Phys. Rev. D}\ }\textbf {\bibinfo {volume} {103}},\ \bibinfo {pages}
  {096021} (\bibinfo {year} {2021})},\ \Eprint
  {http://arxiv.org/abs/2104.11425} {arXiv:2104.11425 [hep-ph]} \BibitemShut
  {NoStop}%
\bibitem [{\citenamefont {Chakraborty}(2019)}]{Chakraborty:2017vvg}%
  \BibitemOpen
  \bibfield  {author} {\bibinfo {author} {\bibfnamefont {P.}~\bibnamefont
  {Chakraborty}},\ }\href {\doibase 10.1140/epjp/i2019-12847-y} {\bibfield
  {journal} {\bibinfo  {journal} {Eur. Phys. J. Plus}\ }\textbf {\bibinfo
  {volume} {134}},\ \bibinfo {pages} {478} (\bibinfo {year} {2019})},\ \Eprint
  {http://arxiv.org/abs/1711.04404} {arXiv:1711.04404 [nucl-th]} \BibitemShut
  {NoStop}%
\end{thebibliography}%
\end{document}